\documentclass[aps,pra,twocolumn,nobibnotes,superscriptaddress]{revtex4}
\usepackage[utf8]{inputenc}
\usepackage[T1]{fontenc}
\usepackage{bm}
\usepackage{ulem}
\usepackage{epsfig}
\usepackage{graphicx}
\usepackage{amssymb,amsmath,amsbsy,amsgen,amsfonts}
\usepackage{dcolumn}
\usepackage{amsthm}
\usepackage{mathtools}
\usepackage{mathrsfs}
\usepackage{latexsym}
\usepackage{array}
\usepackage{color}
\usepackage{amstext}
\usepackage{multirow}
\usepackage{bbold}
\usepackage{color}
\usepackage{xcolor}
\usepackage{listings}
\usepackage{tikz}
\usepackage{lipsum}
\usepackage{textcomp}

\makeatletter
\allowdisplaybreaks[1]

\makeatletter
\let\old@makecaption=\@makecaption
\usepackage{caption}
\usepackage{subcaption}
\let\@makecaption=\old@makecaption
\makeatother

\usepackage{epstopdf} %for texshop on mac

\DeclareSymbolFont{symbols}{OMS}{cmsy}{m}{n}

\newcommand{\appendref}[1]{Appendix~\ref{#1}}

\newcommand{\figref}[1]{Fig.~\ref{#1}}
\newcommand{\tabref}[1]{Table~\ref{#1}}

\definecolor{sciencered}{RGB}{180, 22, 44}
\definecolor{agrigreen}{RGB}{61, 138, 26}
\definecolor{engineeringyellow}{RGB}{255, 156, 47}
\definecolor{edublue}{RGB}{34, 61, 113}

\begin{document}

\title{Quantum random number generation using an on-chip nanowire plasmonic waveguide}

\author{C. Strydom}
\email{conradstryd@gmail.com}
\affiliation{Stellenbosch Photonics Institute, Department of Physics, Stellenbosch University, Stellenbosch, Matieland 7602, South Africa}
\author{S. Soleymani}
\affiliation{Department of Engineering Science and Mechanics, Pennsylvania State University, University Park, State College, PA 16802, USA}
\affiliation{Massachusetts General Hospital --- Harvard Medical School, Boston, MA 02129, USA}
\author{\c{S}. K. \"{O}zdemir}
\affiliation{Department of Engineering Science and Mechanics, Pennsylvania State University, University Park, State College, PA 16802, USA}
\author{M. S. Tame}
\affiliation{Stellenbosch Photonics Institute, Department of Physics, Stellenbosch University, Stellenbosch, Matieland 7602, South Africa}

\begin{abstract}
Quantum random number generators employ the inherent randomness of quantum mechanics to generate truly unpredictable random numbers, which are essential in cryptographic applications.  While a great variety of quantum random number generators have been realised using photonics, few exploit the high-field confinement offered by plasmonics, which enables device footprints an order of magnitude smaller in size.  Here we integrate an on-chip nanowire plasmonic waveguide into an optical time-of-arrival based quantum random number generation setup.  Despite loss, we achieve a random number generation rate of $14.4\,\text{Mbits/s}$ using low light intensity, with the generated bits passing industry standard tests without post-processing.  By increasing the light intensity, we were then able to increase the generation rate to $41.4\,\text{Mbits/s}$, with the resulting bits only requiring a shuffle to pass all tests.  This is an order of magnitude increase in the generation rate and decrease in the device size compared to previous work.  Our experiment demonstrates the successful integration of an on-chip nanoscale plasmonic component into a quantum random number generation setup.  This may lead to new opportunities in compact and scalable quantum random number generation.
\end{abstract}

%\pacs{}

\maketitle

%%%%%%%%%%%%%%%%%%%%%%%%%%%%
%%%%%%%%%%%%%%%%%%%%%%%%%%%%
%%%%%%%%%%%%%%%%%%%%%%%%%%%%
%%%%%%%%%%%%%%%%%%%%%%%%%%%%
\section{Introduction}\label{sec:introduction} 

Random numbers are used extensively in cryptography~\cite{application1}, simulation~\cite{application2} and fundamental physics tests~\cite{application3}, as well as in lotteries, machine learning and coordination in computer networks~\cite{review1}.  When classical techniques are used to generate random numbers, the unpredictability relies on incomplete knowledge, which can result in the random numbers being more predictable than anticipated.  In fact, it was shown in a number of recent studies that even simple machine learning algorithms can successfully predict the random numbers generated by poor quality classical pseudorandom number generators~\cite{PRNGfail1, PRNGfail2, PRNGfail3}, rendering them useless for cryptography.  While high quality cryptographically secure pseudorandom number generators have been developed, they are incredibly resource intensive~\cite{CPRNG1, CPRNG2, CPRNG3}.  Furthermore, it is unclear whether these cryptographically secure classical pseudorandom number generators will be able to withstand future improved machine learning algorithms, since they are still fundamentally deterministic.

On the other hand, when random numbers are generated using quantum mechanical techniques, unpredictability is guaranteed by the inherent randomness of quantum mechanics~\cite{review1, review2, review3}.  A great variety of quantum random number generation schemes have been successfully realised experimentally using photonics, including branching paths~\cite{bp1, bp2, bp3}, time-of-arrival~\cite{toa1, toa2, toa3, toa4, toa5, toa6, toa7, toa8, toa9}, photon counting~\cite{pc1, pc2, pc3}, vacuum fluctuations~\cite{vf1, vf2, vf3, vf4, vf5, vf6} and laser phase fluctuations~\cite{lpf1, lpf2, lpf3, lpf4, lpf5, lpf6}.  More recently, several of these schemes have been realised in the form of on-chip quantum random number generators~\cite{chip1, chip2, chip3, chip4, chip5, chip6, chip7, chip8, chip9}.  Random number generation schemes have also been implemented on cloud-based superconducting quantum computers~\cite{cbs1, cbs2, cbs3}, trapped ions~\cite{ti1} and magnetic tunnel junctions~\cite{mtj1, mtj2}.  Here we investigate quantum random number generation using plasmonics~\cite{plasmonicrev1, plasmonicrev2}.

The past two decades have seen an increasing body of work aimed at understanding quantum features of plasmons and how plasmonic confinement and losses affect the transport of quantum states of light~\cite{qplasmonics1, qplasmonics2, qplasmonics3, qplasmonics4, qplasmonics5, qplasmonics6, qplasmonics7, qplasmonics8, qplasmonics9, qplasmonics10, qplasmonics11, qplasmonics12}.  The ability of plasmonics to confine light to subwavelength scales, well below the diffraction limit~\cite{plasmonicsdl1, plasmonicsdl2}, and to simultaneously carry optical and electrical signals~\cite{plasmonicsoe1, plasmonicsoe2} suggests that quantum information processing protocols can be carried out at a much smaller scale than in the dielectric systems typically used in photonics~\cite{photonics1, photonics2} and that such systems can be integrated with existing electronics via electrically controlled and tuned plasmonic devices via carrier injection and electrical modulation.  Studies have already resulted in an array of applications, such as quantum plasmonic sensing~\cite{qps1, qps2, qps3, qps4, qps5, qps6, qps7, qps8, qps9}, plasmonic entanglement generation~\cite{plasmoniceg1} and distillation~\cite{plasmoniced1, plasmoniced2}, plasmonic quantum gates~\cite{plasmonicqg1} and plasmonic quantum state engineering~\cite{plasmonicqse1}, which have revealed that, despite being inherently lossy, plasmonic components can be successfully employed in quantum information processing tasks.

Recently, the branching paths scheme for quantum random number generation was demonstrated
using an on-chip plasmonic beamsplitter~\cite{plasmonicbs}.  However, other schemes exist, for example the time-of-arrival scheme is very different to the branching paths scheme and has two major advantages, namely it requires one detector instead of two, and substantially higher bit rates are possible as multiple bits of randomness can be extracted from a single photon.  So far, the time-of-arrival scheme has not been demonstrated using plasmonics.  Moreover, most time-of-arrival generators are bulky and use a highly attenuated coherent source as a single-photon source, where photons simply propagate through free space to a single-photon detector~\cite{toa1, toa2, toa3, toa4, toa5, toa6, toa7, toa8, toa9}.  An integrated circuit time-of-arrival generator has been realised recently by adding a semi-coherent silicon LED source directly on a detector chip~\cite{chip2}.  However, an alternative and more flexible option for on-chip time-of-arrival based quantum random number generation is to embed an on-chip source~\cite{source1, source2, source3, source4} and detector~\cite{detector1, detector2, detector3, detector4} within a plasmonic waveguide.

In this paper we report a time-of-arrival based quantum random number generation scheme using a nanowire plasmonic waveguide.  In particular, we integrate an on-chip nanowire plasmonic waveguide into an optical time-of-arrival based quantum random number generation setup and test its performance in the presence of loss.  Despite loss, we initially managed to achieve a random number generation rate of $14.4\,\text{Mbits/s}$.  The generated bits did not require any post-processing to pass the industry standard ENT~\cite{ENT} and NIST~\cite{NIST} Statistical Test Suites.  By increasing the light intensity, we were able to increase the generation rate to $41.4\,\text{Mbits/s}$, however these bits required a shuffle to pass all the tests due to some correlation being introduced.  Our work demonstrates the successful integration of an on-chip nanoscale plasmonic component into a time-of-arrival based quantum random number generation setup.  The specific advantage offered by introducing an on-chip nanowire plasmonic waveguide into the setup is that light in the plasmonic waveguide can be confined to a length scale well below the diffraction limit, which enables the footprint of the on-chip component to be reduced to a size unattainable with dielectric hardware~\cite{plasmonicsdl1, plasmonicsdl2}.  This research therefore makes an important contribution to addressing the on-going challenge of miniaturising on-chip quantum random number generators, as it shows how an on-chip nanoscale plasmonic component, with a footprint well below that of equivalent state-of-the-art dielectric components, can be successfully used in quantum random number generation despite it being inherently lossy.  Although our current setup employs an off-chip source and detector, future work on the integration of an on-chip source~\cite{source1, source2, source3, source4} and detector~\cite{detector1, detector2, detector3, detector4} will yield a fully integrated nanophotonic quantum random number generator chip with a footprint an order of magnitude smaller than its dielectric counterpart.  This opens up new opportunities in compact and scalable quantum random number generation.

%%%%%%%%%%%%%%%%%%%%%%%%%%%%
%%%%%%%%%%%%%%%%%%%%%%%%%%%%
%%%%%%%%%%%%%%%%%%%%%%%%%%%%
%%%%%%%%%%%%%%%%%%%%%%%%%%%%
\section{Experimental setup}\label{sec:setup} 

\begin{figure*}
    \centering
    \begin{subfigure}[b]{.69\textwidth}
        \centering
        \begin{tikzpicture}[scale=0.35, font=\scriptsize]
        \draw[edublue!50, fill=edublue!20] (13.5, 0) rectangle (15, -1.5);
        \draw[edublue!50] (13.5, -1.5) -- (15, 0);
        \draw[sciencered, line width=0.25mm] (2, -0.75) -- (26, -0.75);
        \draw[gray, fill=gray!65] (0, -0.5) -- (0, -1) -- (0.5, -1.5) -- (2, -1.5) -- (2, 0) -- (0.5, 0) -- cycle;
        \draw[agrigreen, fill=agrigreen] (1.2, -0.04) rectangle (1.3, -1.46);
        \draw[agrigreen, fill=agrigreen] (1.5, -0.04) rectangle (1.6, -1.46);
        \draw[edublue!75, fill=edublue!75] (-0.1, -1) -- (-0.1, -0.5) -- (-0.7, -0.5) arc (90:270:0.25) -- cycle;
        \draw[edublue!75] plot [smooth, tension=2] coordinates {(-0.95, -0.75) (-1.95, -0.75) (-2.05, -0.25) (-3, -0.25)};
        \draw[gray, fill=gray!65] (4.5, 0.25) rectangle (5, -1.75);
        \draw[sciencered, fill=sciencered!50] (7.5, 0) rectangle (8, -1.5);
        \draw[engineeringyellow, fill=engineeringyellow!50] (10.5, 0) rectangle (11, -1.5);
        \draw[sciencered, fill=sciencered!50] (17.5, 0) rectangle (18, -1.5);
        \draw[black!80, fill=black] (26, 0) -- (26, -1.5) -- (27.5, -1.5) -- (28, -1) -- (28, -0.5) -- (27.5, 0) -- cycle;
        \draw[engineeringyellow, fill=engineeringyellow] (26.4, -0.04) rectangle (26.5, -1.46);
        \draw[engineeringyellow, fill=engineeringyellow] (26.7, -0.04) rectangle (26.8, -1.46);
        \draw[edublue!75, fill=edublue!75] (28.25, 0.5) rectangle (28.35, -2);
        \draw[gray, fill=gray!65] (20.5, 0.5) -- (22, 0.5) -- (22, 2) -- (21.5, 2.5) -- (21, 2.5) -- (20.5, 2) -- cycle;
        \draw[agrigreen, fill=agrigreen] (20.54, 0.9) rectangle (21.96, 1);
        \draw[agrigreen, fill=agrigreen] (20.54, 1.2) rectangle (21.96, 1.3);
        \draw[engineeringyellow, fill=engineeringyellow] (21, 2.6) -- (21.5, 2.6) -- (21.5, 3.2) arc (0:180:0.25) -- cycle;
        \draw[engineeringyellow] plot [smooth, tension=0.5] coordinates {(21.25, 3.45) (21, 3.8) (20, 4) (19.75, 4.35)};
        \draw[agrigreen, line width=0.25mm] (21.25, 0.5) -- (21.25, -1.15) -- (26, -1);
        \draw[engineeringyellow, fill=engineeringyellow] (21, -0.9) -- (21.5, -1.4) -- (21.4, -1.5) -- (20.9, -1) -- cycle;
        \node[] at (-2.6, 1.2) {from};
        \node[] at (-2.6, 0.6) {Source};
        \node[] at (-1.3, -1.6) {SM};
        \node[] at (1.25, -2.25) {20x};
        \node[] at (1.25, 0.75) {BE};
        \node[] at (4.75, -2.75) {NDF};
        \node[] at (7.75, -2.25) {HWP};
        \node[] at (10.75, -2.25) {QWP};
        \node[] at (14.25, -2.25) {PBS};
        \node[] at (17.75, -2.25) {HWP};
        \node[] at (26.75, -2.25) {100x};
        \node[] at (26.75, -3.05) {DLM};
        \node[] at (28.3, 2.8) {Nanowire};
        \node[] at (28.3, 2) {Plasmonic};
        \node[] at (28.3, 1.2) {Waveguide};
        \node[] at (21.25, -2.25) {KM};
        \node[] at (22.9, 1.25) {20x};
        \node[] at (19.6, 1.25) {FC};
        \node[] at (22.2, 3.9) {MM};
        \node[] at (18.4, 4.2) {to};
        \node[] at (18.4, 3.6) {Detector};
        \end{tikzpicture}
        \vspace{-0.1cm}
        \caption{Experimental Setup}
        \label{fig:setup}
    \end{subfigure}
    \begin{subfigure}[b]{.29\textwidth}
        \centering
        \includegraphics[scale=0.5]{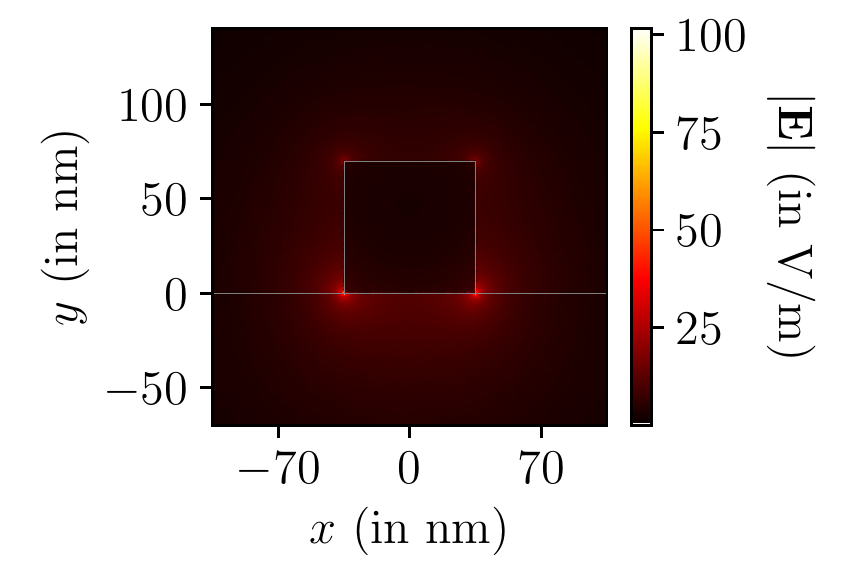}
        \addtocounter{subfigure}{1}
        \vspace{-0.1cm}
        \caption{Waveguide Mode}
        \label{fig:mode}
    \end{subfigure}
    \begin{subfigure}[b]{.49\textwidth}
        \centering
        \vspace{0.5cm}
        \begin{tikzpicture}[scale=0.3, font=\scriptsize]
        \draw[engineeringyellow, fill=engineeringyellow] (0, 0) rectangle (0.25, 2);
        \draw[engineeringyellow, fill=engineeringyellow] (0.6, 0) rectangle (0.99, 2);
        \draw[engineeringyellow, fill=engineeringyellow] (1.34, 0) rectangle (1.73, 2);
        \draw[engineeringyellow, fill=engineeringyellow] (2.08, 0) rectangle (2.47, 2);
        \draw[engineeringyellow, fill=engineeringyellow] (2.82, 0) rectangle (3.21, 2);
        \draw[engineeringyellow, fill=engineeringyellow] (3.56, 0) rectangle (3.95, 2);
        \draw[engineeringyellow, fill=engineeringyellow] (4.3, 0) rectangle (4.69, 2);
        \draw[engineeringyellow, fill=engineeringyellow] (5.04, 0) rectangle (5.43, 2);
        \draw[engineeringyellow, fill=engineeringyellow] (5.78, 0) rectangle (6.17, 2);
        \draw[engineeringyellow, fill=engineeringyellow] (6.52, 0) rectangle (6.91, 2);
        \draw[engineeringyellow, fill=engineeringyellow] (7.26, 0) rectangle (7.65, 2);
        \draw[engineeringyellow, fill=engineeringyellow] (8, 0) -- (8, 2) -- (9.266, 2) -- (11.673, 1.035) -- (15.02, 1.035) -- (17.427, 2) -- (18.85, 2) -- (18.85, 0) -- (17.427, 0) -- (15.02, 0.965) -- (11.673, 0.965) -- (9.266, 0) -- cycle;
        \draw[engineeringyellow, fill=engineeringyellow] (19.2, 0) rectangle (19.59, 2);
        \draw[engineeringyellow, fill=engineeringyellow] (19.94, 0) rectangle (20.33, 2);
        \draw[engineeringyellow, fill=engineeringyellow] (20.68, 0) rectangle (21.07, 2);
        \draw[engineeringyellow, fill=engineeringyellow] (21.42, 0) rectangle (21.81, 2);
        \draw[engineeringyellow, fill=engineeringyellow] (22.16, 0) rectangle (22.55, 2);
        \draw[engineeringyellow, fill=engineeringyellow] (22.9, 0) rectangle (23.29, 2);
        \draw[engineeringyellow, fill=engineeringyellow] (23.64, 0) rectangle (24.03, 2);
        \draw[engineeringyellow, fill=engineeringyellow] (24.38, 0) rectangle (24.77, 2);
        \draw[engineeringyellow, fill=engineeringyellow] (25.12, 0) rectangle (25.51, 2);
        \draw[engineeringyellow, fill=engineeringyellow] (25.86, 0) rectangle (26.25, 2);
        \draw[engineeringyellow, fill=engineeringyellow] (26.6, 0) rectangle (26.86, 2);
        \node[] at (3.75, -1) {input (light)};
        \node[] at (13.35, -1) {SPP};
        \node[] at (23.1, -1) {output (light)};
        \end{tikzpicture}
        \addtocounter{subfigure}{-2}
        \caption{Nanowire Plasmonic Waveguide}
        \label{fig:waveguide}
    \end{subfigure}
    \begin{subfigure}[b]{.49\textwidth}
        \centering
        \vspace{0.2cm}
        \begin{tikzpicture}[scale=0.8, font=\scriptsize]
        \draw[very thick] (0, 0) -- (6, 0);
        \draw[very thick] (0, -0.3) -- (0, 0.3);
        \draw[very thick] (1, -0.3) -- (1, 0.3);
        \draw[very thick] (2, -0.3) -- (2, 0.3);
        \draw[very thick] (3, -0.3) -- (3, 0.3);
        \draw[very thick] (4, -0.3) -- (4, 0.3);
        \draw[very thick] (5, -0.3) -- (5, 0.3);
        \draw[very thick] (6, -0.3) -- (6, 0.3);
        \draw (0, 0.8) -- (0.6, 0.8) -- (0.6, 1.2) -- (0.8, 1.2) -- (0.8, 0.8) -- (3.4, 0.8) -- (3.4, 1.2) -- (3.6, 1.2) -- (3.6, 0.8) -- (4.7, 0.8) -- (4.7, 1.2) -- (4.9, 1.2) -- (4.9, 0.8) -- (6, 0.8);
        \draw[dashed] (0.6, 0) -- (0.6, 0.8);
        \draw[dashed] (3.4, 0) -- (3.4, 0.8);
        \draw[dashed] (4.7, 0) -- (4.7, 0.8);
        \draw[latex-latex] (0, 0.3) -- (0.6, 0.3);
        \draw[latex-latex] (3, 0.3) -- (3.4, 0.3);
        \draw[latex-latex] (4, 0.3) -- (4.7, 0.3);
        \node[] at (0.5, -0.18) {$T$};
        \node[] at (1.5, -0.18) {$T$};
        \node[] at (2.5, -0.18) {$T$};
        \node[] at (3.5, -0.18) {$T$};
        \node[] at (4.5, -0.18) {$T$};
        \node[] at (5.5, -0.18) {$T$};
        \node[] at (0.4, 1.1) {$t_1$};
        \node[] at (3.2, 1.1) {$t_2$};
        \node[] at (4.5, 1.1) {$t_3$};
        \node[] at (6.9, 0) {Reference};
        \node[] at (7.15, 0.9) {Arrival times};
        \end{tikzpicture}
        \addtocounter{subfigure}{1}
        \caption{Time-of-arrival Scheme}
        \label{fig:scheme}
    \end{subfigure}
    \begin{subfigure}[b]{.98\textwidth}
        \centering
        \vspace{0.3cm}
        \begin{tikzpicture}[font=\scriptsize]
        \node[] at (0, 0) {\includegraphics[trim=2.2cm 0cm 4.6cm 2.3cm, clip, scale=0.45]{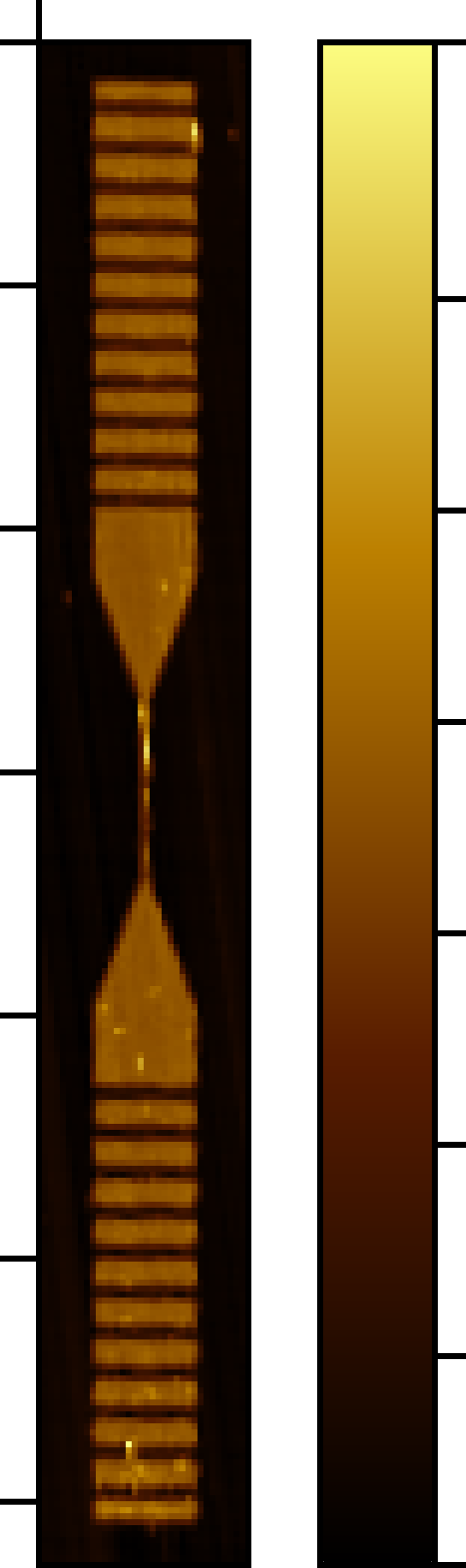}};
        \node[] at (-1.43, 3.97) {0};
        \node[] at (-1.43, 2.67) {5};
        \node[] at (-1.51, 1.37) {10};
        \node[] at (-1.51, 0.07) {15};
        \node[] at (-1.51, -1.23) {20};
        \node[] at (-1.51, -2.53) {25};
        \node[] at (-1.51, -3.83) {30};
        \node[] at (-1.04, 4.37) {0};
        \node[] at (-0.52, 4.37) {2};
        \node[] at (0, 4.37) {4};
        \draw[thick] (-0.52, 4.19) -- (-0.52, 3.98);
        \draw[thick] (0, 4.19) -- (0, 3.98);
        \node[] at (1.46, -4.18) {0};
        \node[] at (1.54, -3.05) {20};
        \node[] at (1.54, -1.92) {40};
        \node[] at (1.54, -0.79) {60};
        \node[] at (1.54, 0.34) {80};
        \node[] at (1.62, 1.47) {100};
        \node[] at (1.62, 2.6) {120};
        \node[] at (1.62, 3.97) {144};
        \node[] at (-0.52, 4.8) {$x$ (in $\mu$m)};
        \node[rotate=90] at (-1.94, 0.07) {$y$ (in $\mu$m)};
        \node[] at (1.3, 4.4) {Height (in nm)};
        \end{tikzpicture}
        \begin{tikzpicture}[font=\scriptsize]
        \node[] at (0, 0) {\includegraphics[trim=0.78cm 0cm 2.27cm 1.17cm, clip, scale=0.45]{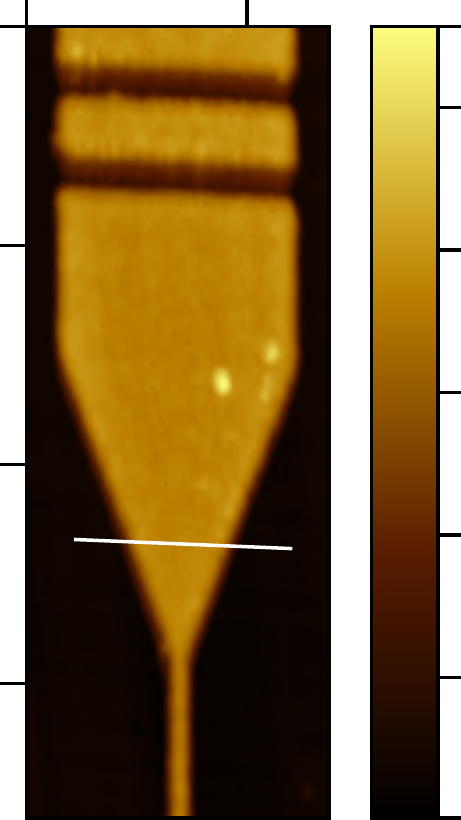}};
        \node[] at (-1.2, 1.75) {0};
        \node[] at (-1.2, 0.745) {2};
        \node[] at (-1.2, -0.26) {4};
        \node[] at (-1.2, -1.265) {6};
        \node[] at (-0.93, 2.05) {0};
        \node[] at (0.075, 2.05) {2};
        \node[] at (1.22, -1.86) {0};
        \node[] at (1.3, -1.215) {20};
        \node[] at (1.3, -0.57) {40};
        \node[] at (1.3, 0.075) {60};
        \node[] at (1.3, 0.72) {80};
        \node[] at (1.38, 1.365) {100};
        \node[] at (1.38, 1.75) {111};
        \node[] at (-0.3, 2.45) {$x$ (in $\mu$m)};
        \node[rotate=90] at (-1.6, 0.24) {$y$ (in $\mu$m)};
        \node[] at (1.48, 2.15) {Height (in nm)};
        \node[] at (0.7, -4.3) {\includegraphics[trim=0cm 0.3cm 0cm 0cm, clip, scale=0.5]{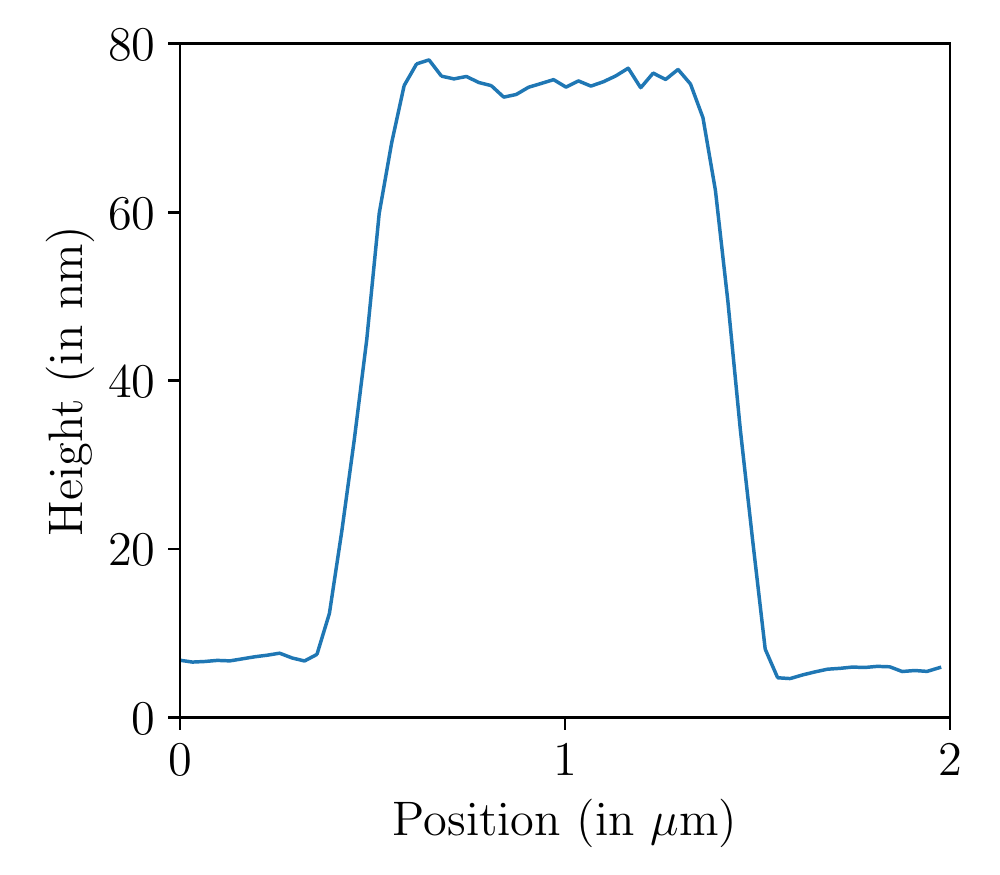}};
        \end{tikzpicture}
        \begin{tikzpicture}[font=\scriptsize]
        \node[] at (0, 0) {\includegraphics[trim=0.81cm 0cm 2.41cm 1.19cm, clip, scale=0.45]{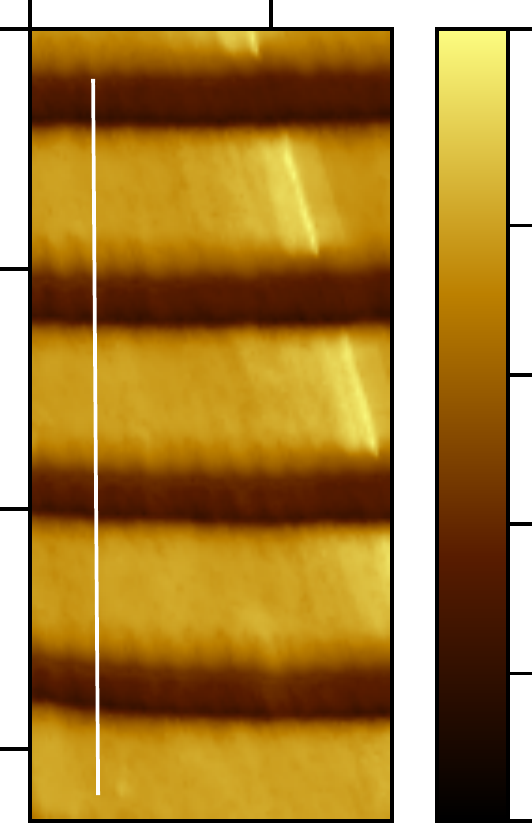}};
        \node[] at (-1.365, 1.75) {0};
        \node[] at (-1.365, 0.655) {1};
        \node[] at (-1.365, -0.44) {2};
        \node[] at (-1.365, -1.535) {3};
        \node[] at (-1.085, 2.05) {0};
        \node[] at (0.01, 2.05) {1};
        \node[] at (1.365, -1.86) {0};
        \node[] at (1.445, -1.18) {20};
        \node[] at (1.445, -0.5) {40};
        \node[] at (1.445, 0.18) {60};
        \node[] at (1.445, 0.86) {80};
        \node[] at (1.525, 1.75) {106};
        \node[] at (-0.355, 2.45) {$x$ (in $\mu$m)};
        \node[rotate=90] at (-1.765, 0.107) {$y$ (in $\mu$m)};
        \node[] at (1.625, 2.15) {Height (in nm)};
        \node[] at (0.55, -4.3) {\includegraphics[trim=0cm 0.3cm 0cm 0cm, clip, scale=0.5]{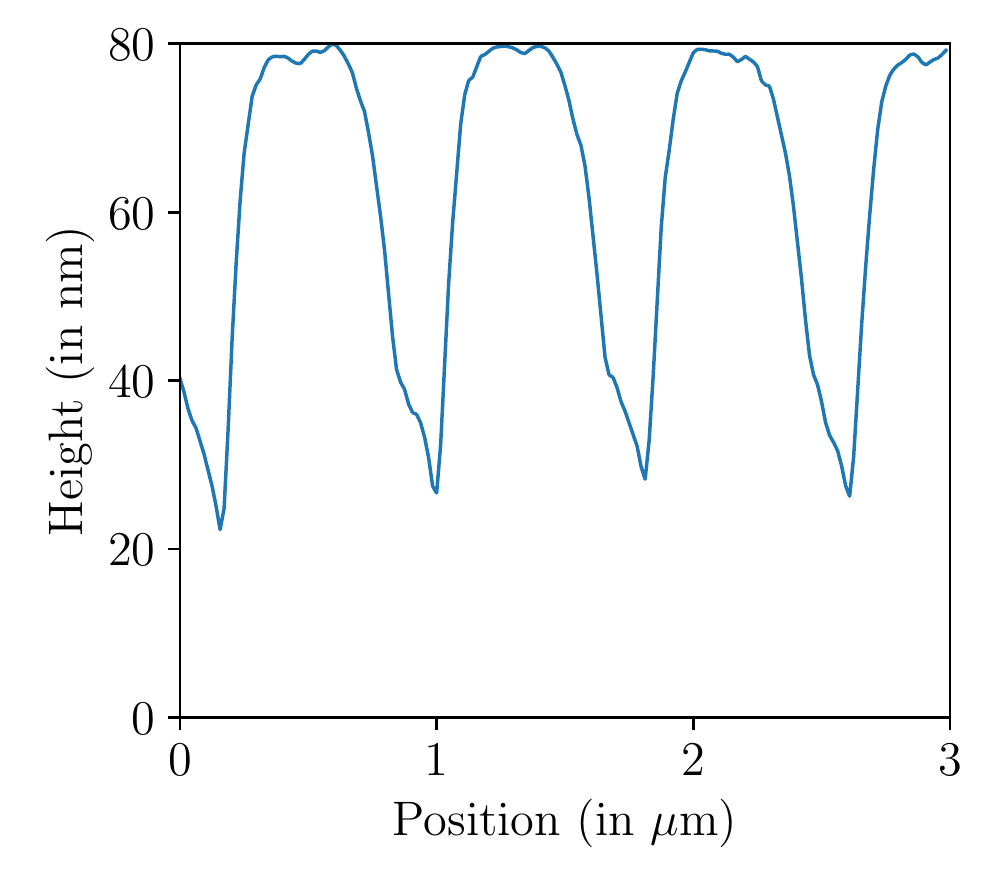}};
        \end{tikzpicture}
        \caption{Atomic Force Microscope Images}
        \label{fig:afm}
    \end{subfigure}
    \caption{Quantum random number generation using an on-chip nanowire plasmonic waveguide.  (a)~Experimental Setup shows the setup used to investigate time-of-arrival based quantum random number generation using an on-chip nanowire plasmonic waveguide.  The labels used are: single-mode optical fibre (SM), beam expander (BE), neutral density filter (NDF), half-wave plate (HWP), quarter-wave plate (QWP), polarising beamsplitter (PBS), diffraction-limited microscope (DLM), knife-edge mirror (KM), fibre coupler (FC) and multi-mode optical fibre (MM).  (b)~Nanowire Plasmonic Waveguide shows a top view of the on-chip nanowire plasmonic waveguide used in the experiments.  (c)~Waveguide Mode shows the electric field distribution of the characteristic mode through a cross-section of the nanowire waveguide for a vacuum wavelength of $785\,\text{nm}$.  The grey square shows the nanowire waveguide and the grey line shows the substrate surface.  (d)~Time-of-arrival Scheme illustrates the implemented variation of the time-of-arrival scheme, in which random numbers are obtained from the arrival times of photons relative to an external time reference~\cite{toa6}.  (e)~Atomic Force Microscope Images shows atomic force microscope (AFM) images of the fabricated on-chip nanowire plasmonic waveguide.  These include an AFM image of the entire nanowire plasmonic waveguide (left), an AFM image of the top tapering (top centre), an AFM height profile of the top tapering (bottom centre), an AFM image of the top grating (top right) and an AFM height profile of the top grating (bottom right).}
    \label{fig:QRNGnpw}
\end{figure*}
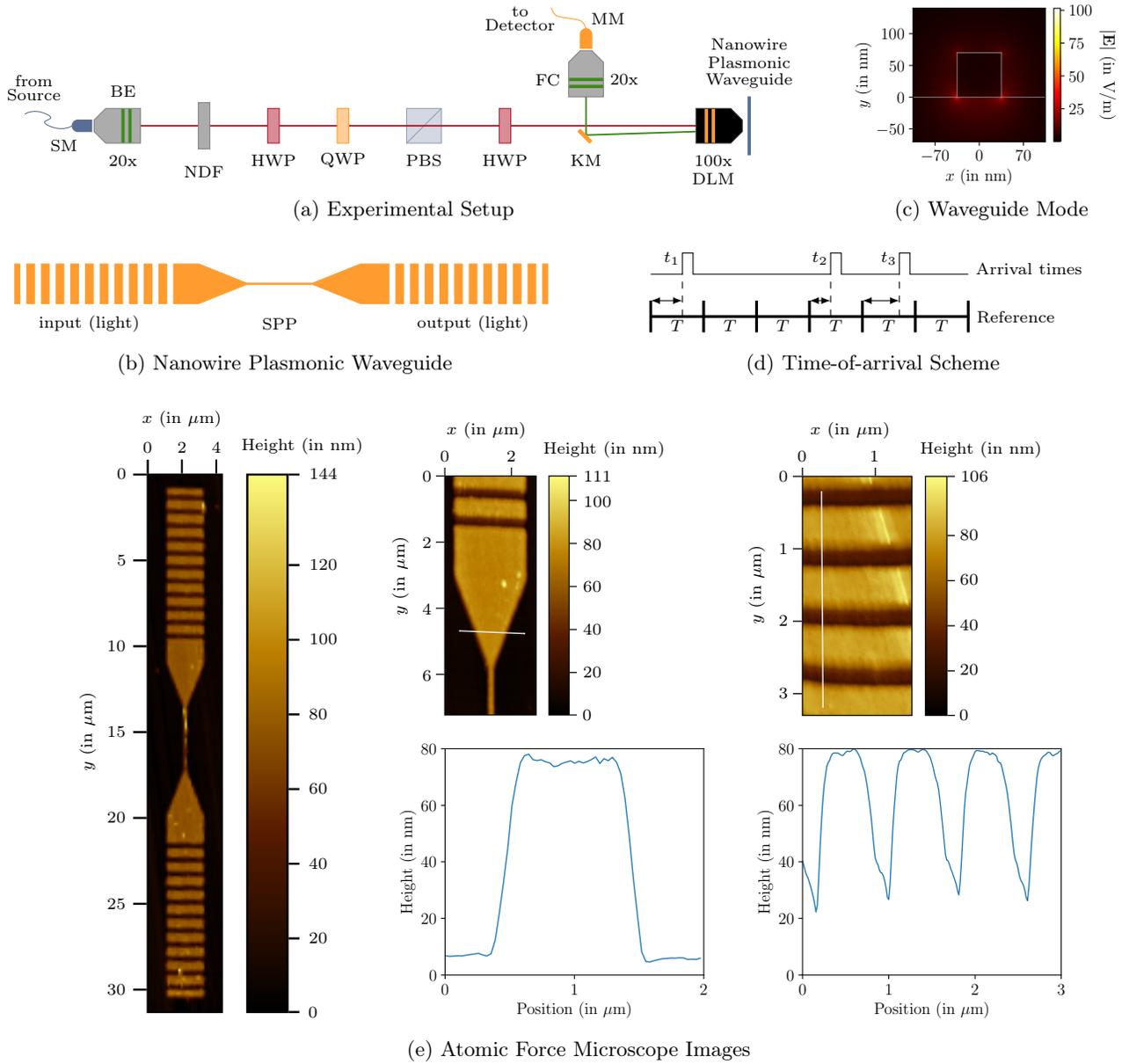

The experimental setup used to investigate time-of-arrival based quantum random number generation using a nanowire plasmonic waveguide is shown in \figref{fig:setup}.  The plasmonic waveguide used in the experiments comprises a gold nanowire $70\,\text{nm}$ in diameter and just over $3\,\mu\text{m}$ in length with tapering and a grating with a period of $740\,\text{nm}$ on either end (see \figref{fig:waveguide}).  Each 11-step grating is $2\,\mu\text{m}$ in width and $70\,\text{nm}$ in height.  In the optical setup, highly attenuated coherent laser light, with a vacuum wavelength of $785\,\text{nm}$, is focused onto the input grating of the nanowire plasmonic waveguide using a diffraction-limited microscope (DLM) objective.  At the input grating, photons are converted to surface plasmon polaritons, which propagate through the tapering regions and the nanowire waveguide to the output grating, where they are converted back to photons.  The gratings provide free-space photons with the additional momentum needed to couple into the bound plasmonic waveguide mode, thereby enabling the conversion of photons to surface plasmon polaritons and vice versa~\cite{plasmonicbs, plasmonicsthesis, gratings, polarisation2}.  The tapering regions adiabatically nanofocus surface plasmon polaritons into and out of the nanowire waveguide~\cite{taperings}.  The effective mode index of the characteristic mode of the gold nanowire waveguide was determined to be $n_{\text{eff}}=1.84+0.0573\text{i}$ using a 2D finite element method simulation in COMSOL (see \appendref{append:simulation}).  The associated electric field distribution through a cross-section of the nanowire waveguide is shown in \figref{fig:mode}.  We see that the electric field is highly confined at the corners of the nanowire waveguide, at a length scale well below the diffraction limit --- something which would have been impossible with a dielectric waveguide.  Photons are collected from the output grating of the nanowire plasmonic waveguide using the same DLM objective that was used to focus light onto its input grating, and are then sent to a single-photon detector.  The arrival times of photons at the detector, relative to an external reference (see \figref{fig:scheme}), are then used to obtain random numbers~\cite{toa6}, as will be explained in more detail later.

The temporal degree of freedom of photons generated during stimulated emission is a true source of randomness~\cite{toa8}.  In this work we employ a $\lambda=785\,\text{nm}$ continuous-wave laser (Thorlabs LPS-785-FC) operating in the stimulated emission regime.  The wavelength of our continuous-wave laser source is chosen to be well within the coupling window of the $740\,\text{nm}$-period gratings of our on-chip nanowire plasmonic waveguide, which ensures that the conversion between photons and surface plasmon polaritons takes place with maximal efficiency at these gratings~\cite{plasmonicbs, plasmonicsthesis}.  A polarisation-preserving single-mode optical fibre (SM) connects the continuous-wave laser to a beam expander (BE), through which polarised coherent laser light enters the optical setup in \figref{fig:setup}.  The collimated beam passes through a neutral density filter (NDF), a half-wave plate (HWP), a quarter-wave plate (QWP), a polarising beamsplitter (PBS) and a second HWP.  The NDF, along with loss in the optical setup and the plasmonic waveguide, ensures that light reaching the detector is attenuated down to an appropriate level for single-photon detection.  The first HWP, the QWP and the PBS are used to purify the polarisation of light from the laser.  In particular, the PBS transmits only horizontally polarised photons and the preceding HWP and QWP adjust the polarisation of light incident on the PBS so as to maximise the transmission of light through the PBS.  The second HWP rotates the resulting horizontally polarised light so as to maximise the conversion of photons to surface plasmon polaritons at the input grating of the nanowire plasmonic waveguide, which can only be achieved when the polarisation vector is perpendicular to the gratings~\cite{plasmonicbs, plasmonicsthesis}.  A 100x DLM objective is used to focus the beam onto the input grating of the nanowire plasmonic waveguide at a spot size of about $2\,\mu\text{m}$.

The plasmonic waveguide is fabricated on a silica glass substrate with a refractive index of $1.5255$ and a thickness of $1\,\text{mm}$ using a combination of electron beam lithography and electron beam deposition.  A resist is first spin coated on the silica glass substrate and $20\,\text{nm}$ of gold is deposited so that the surface becomes conductive.  Electron beam lithography is used to define the regions for the nanowire, the taperings and the gratings.  The gold is then etched and resist is developed.  Next a lift-off technique is employed, where first a $5\,\text{nm}$ thick adhesion layer of titanium is deposited and then the desired $70\,\text{nm}$ thick gold layer on the silica glass substrate.  The unexposed resist and gold is then lifted off with acetone, IPA and de-ionised water.  Atomic force microscope (NT-MDT Smena) images of the fabricated on-chip nanowire plasmonic waveguide are shown in \figref{fig:afm}.

The power transmission factor of the nanowire plasmonic waveguide was measured to be $\eta_{\text{wgd}}=6.7\times 10^{-5}$ (see \appendref{append:loss}).  Losses occur in the waveguide as a result of scattering during the conversion between photons and surface plasmon polaritons at the gratings, leakage from the tapering regions, and absorption during the propagation of surface plasmon polaritons along the nanowire.  By using results from \appendref{append:simulation}, we were able to estimate the power transmission factor of the nanowire as well as the net power transmission factor of a grating and tapering region in \appendref{append:loss}.  Hence Appendices~\ref{append:simulation} and~\ref{append:loss} play a very important role in the characterisation of our main component by providing us with a graded picture of the different parts of our on-chip nanowire plasmonic waveguide and a quantitative analysis of the losses occurring in each of these parts, which will be of interest to researchers looking to investigate further optimisations and extensions.  Note that one can use higher intensity light to keep the photon detection rate within a certain desired range even in the presence of such losses.  Unlike in a previously demonstrated plasmonic quantum random number generator~\cite{plasmonicbs}, there is no limit on the amount by which one can increase the intensity of the coherent source, since the time-of-arrival scheme does not require the nanowire plasmonic waveguide to operate in the single-excitation regime~\cite{toa6}.

Photons are collected from the output grating of the nanowire plasmonic waveguide by the same DLM objective that was used to focus the input beam onto the input grating.  A knife-edge mirror (KM) is then used to reflect these photons into a fibre coupler (FC), which is connected to a single-photon avalanche diode (SPAD) detector (Excelitas SPCM-AQRH-15) by a multi-mode optical fibre (MM).  The polarisation dependence of the photon detection rate~\cite{polarisation1, polarisation2} confirms that the collection optics is indeed capturing out-coupled photons from the output grating and not scattered photons from the input beam (see \appendref{append:polarisation}).  The SPAD detector used in the experiments has a dead time of $24\,\text{ns}$, a timing resolution of $350\,\text{ps}$ and a maximum dark count rate of $50\,\text{counts/s}$.  For data collection, the SPAD detector is connected to a channel of a Picoquant TimeHarp 260, which is connected to a PC.  The TimeHarp is capable of recording the arrival time of a photon at the detector to a precision of $25\,\text{ps}$.

We implement a variation of the time-of-arrival scheme, first proposed by Nie \textit{et al.}~\cite{toa6}, in which random numbers are obtained from the arrival times of photons relative to an external reference (see \figref{fig:scheme}).  The benefit of this variation compared to earlier versions~\cite{toa1, toa4, toa5}, in which random numbers are obtained directly from the difference of consecutive photon arrival times, is a significant reduction in bias.  In the variation proposed by Nie \textit{et al.}~\cite{toa6}, the external time reference is divided into time intervals of an arbitrary but fixed length $T$, each of which are in turn divided into $N$ bins of equal width.  Provided that at most one photon detection can occur in a time interval of length $T$, it follows that a photon detection occurring in a time interval of length $T$ occurs in each of the $N$ bins with probability $\frac{1}{N}$ (shown in \appendref{append:proof}).  Hence the numbers of the bins in which photon detections occur are uniformly distributed random $\log_2(N)$-bit unsigned integers.  The model proposed by Nie \textit{et al.}~\cite{toa6} for the physical system used in their experimental implementation also applies to our experimental setup and we make similar assumptions and approximations.  Device imperfections which could result in deviations from uniformity and degrade the quality of the random numbers generated by the system are discussed in detail in Refs.~\cite{toa6, toa8}.  These include a non-unit SPAD detector efficiency, SPAD detector dark counts, SPAD detector dead time and a non-zero probability of multi-photon emission from the attenuated coherent source.

For our experiment, we set $T=12.8\,\text{ns}$.  This ensures that $T$ is less than the dead time of the SPAD detector, which in turn ensures that at most one photon detection can occur in a time interval of length $T$.  Furthermore, we set $N=2^8=256$, which enables us to extract a random $8$-bit unsigned integer from each photon arrival time.  We note that the bin width, $\frac{T}{N}=50\,\text{ps}$, is greater than the precision of the TimeHarp, but less than the timing resolution of the SPAD detector.  However, results from previous experiments~\cite{toa6, toa8} suggest that the timing resolution of the detector does not significantly affect the uniformity or the quality of the random numbers.

For data collection, we adjust the light intensity so as to give a photon detection rate of $1.8\,\text{Mcounts/s}$.  This corresponds to an average time interval of $0.56\,\mu\text{s}$ between photon detections, which is much greater than the dead time of the SPAD detector, ensuring that the majority of photons arriving at the SPAD detector are indeed detected.  Random 8-bit unsigned integers extracted from recorded photon arrival times are converted to binary form, resulting in a sample of binary digits or bits.  The most conservative information-theoretic measure with which to quantify the randomness or unpredictability of the bits generated by our setup is the min-entropy~\cite{toa6, toa8}.  As part of the model of the physical system used in their experiment, Nie \textit{et al.}~\cite{toa6} show that the min-entropy of bits generated by a non-ideal experimental implementation of their time-of-arrival scheme can be estimated using the mean number of photons in a time interval of length $T$, which can be calculated using the photon detection rate.  In \appendref{append:entropy}, we show that for our chosen photon detection rate of $1.8\,\text{Mcounts/s}$, the mean photon number is $0.024$, which gives a min-entropy of $0.998$ per bit.  This is very close to the information-theoretically optimal value of $1$ per bit, which shows that our experimental setup is capable of creating very high-quality randomness even in the presence of the device imperfections considered in the model used by Nie \textit{et al.}~\cite{toa6}.

In their own implementation, Nie \textit{et al.}~\cite{toa6} operated their SPAD detector at its saturation rate in order to maximise the random number generation rate.  Consequently, the min-entropy of the raw bits generated by their setup was only $0.88$ per bit and a randomness extractor was needed to increase the min-entropy.  In contrast, the min-entropy for our experimental setup is very close to the information-theoretically optimal value of $1$ per bit, even with device imperfections which can degrade the quality of the randomness taken into account, and so randomness extraction is not required here.  We generated a sample of $866,893,768$ bits in $60\,\text{s}$, which corresponds to a random number generation rate of about $14.4\,\text{Mbits/s}$.  Note that a previous plasmonic quantum random number generator was only able to achieve a generation rate of $2.37\,\text{Mbits/s}$~\cite{plasmonicbs} and a previous on-chip time-of-arrival generator was only able to achieve a generation rate of $1\,\text{Mbits/s}$~\cite{chip2}.  Hence our work demonstrates an order of magnitude improvement in speed compared to these previous devices.  In the next section, we apply a number of industry standard tests~\cite{ENT, NIST} to the first $800\,\text{Mbits}$ generated, which we will refer to as the generated sample.

%%%%%%%%%%%%%%%%%%%%%%%%%%%%
%%%%%%%%%%%%%%%%%%%%%%%%%%%%
%%%%%%%%%%%%%%%%%%%%%%%%%%%%
%%%%%%%%%%%%%%%%%%%%%%%%%%%%
\section{Results}\label{sec:results} 

\begin{figure}
    \centering
    \includegraphics[scale=0.5]{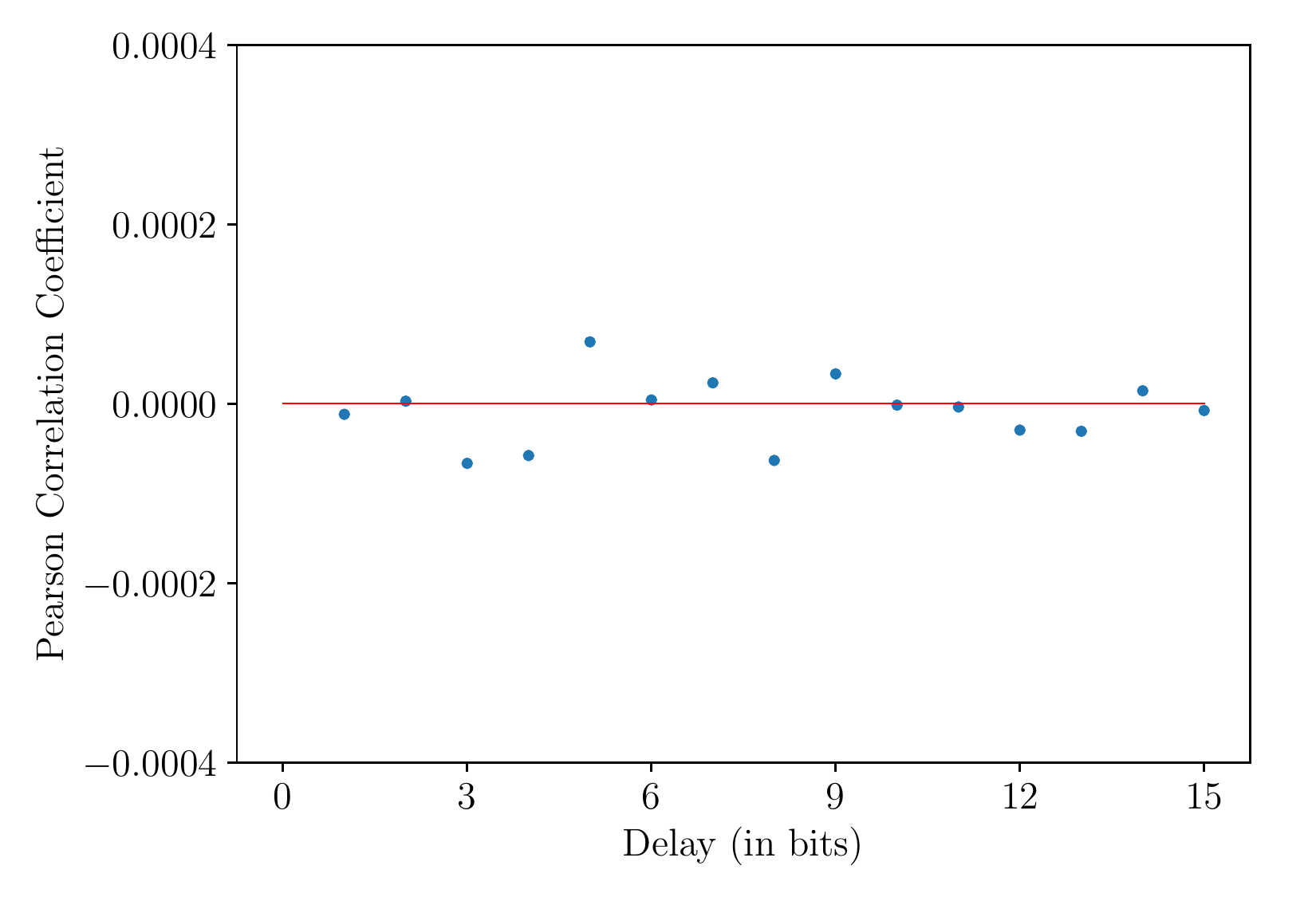}
    \caption{Pearson correlation coefficient of the generated sample with $1$-bit to $15$-bit delays of itself.}
    \label{fig:pearsongen}
\end{figure}

As a first test, we employ the Pearson correlation coefficient~\cite{Pearson} to detect short-ranged correlations in the generated sample.  The Pearson correlation coefficient is a real number in the interval $[-1, 1]$, where a positive value suggests a positive correlation, a negative value suggests a negative correlation and a value close to zero suggests no correlation.  As can be seen in \figref{fig:pearsongen}, short-ranged correlations in the generated sample are negligible.  Furthermore, the relative frequency of zeros and ones in the generated sample is 0.49995 and 0.50005 respectively.  Hence the bias in the generated sample is also negligible.

For further quality assessment, we apply the ENT Statistical Test Suite~\cite{ENT} to the generated sample.  The ENT Statistical Test Suite comprises five simple tests in which the sample being tested is used to compute five important values.  These include the entropy per byte, which quantifies the information density of the sample, the $\chi^2$ distribution, which is known to be extremely sensitive to flaws in random number generators, the arithmetic mean of $8$-bit unsigned integers extracted from the sample, which aids in detecting bias, the Monte Carlo value for $\pi$, which provides a practical test of the generator's suitability for use in simulation, and the serial correlation coefficient, which quantifies correlations between adjacent bytes in the sample.  The quality of the sample being tested can then be assessed by comparing the values obtained using the sample to the known values for a true random sample.  The ENT Statistical Test Suite results for the generated sample are given in \tabref{tab:ENTgen}.  For all five tests, the values obtained using the generated sample show good agreement with the expected values for a true random sample.

\begin{table}[]
\begin{tabular}{|l|l|l|}
\hline
\textbf{Test}                  & \textbf{Generated}        & \textbf{Expected}           \\ \hline
Entropy                        & 7.999998                  & 8.000000                    \\ \hline
$\chi^2$ Distribution          & 9.08\%                    & 10--90\%                    \\ \hline
Arithmetic Mean                & 127.503                   & 127.500                     \\ \hline
Monte Carlo value for $\pi$    & 3.14177173                & 3.14159265                  \\ \hline
Serial Correlation Coefficient & 0.001500                  & 0.000000                    \\ \hline
\end{tabular}
\caption{ENT Statistical Test Suite results for the generated sample.  `Generated' shows the values obtained using the generated sample.  `Expected' shows the expected values for a true random sample.}
\label{tab:ENTgen}
\end{table}

As a final assessment of the quality, we apply the NIST Statistical Test Suite~\cite{NIST} to the generated sample.  The NIST Statistical Test Suite consists of 15 stringent tests, which are primarily aimed at assessing a random number generator's suitability for use in cryptographic applications.  To apply one of these tests to a sample of bits from a generator, the sample is first divided into sequences of a fixed length.  The test is then applied to each sequence and a $p$-value, which can be used to assess the uniformity of the distribution of the test results obtained for the individual sequences, is determined.  For the sample to pass the test, a sufficient number of sequences must pass the test and the $p$-value must be greater than or equal to $0.0001$.  The NIST Statistical Test Suite results for the generated sample are shown in \tabref{tab:NISTgen}.  For each test, the generated sample was divided into 800 sequences $1\,\text{Mbit}$ in length.  The default values were used for the block length, except in the Block Frequency test, where the block length was adjusted from $2^{7}=128$ to $2^{14}=16384$.  The generated sample passed all 15 NIST tests.  Hence we conclude that the random numbers generated by our plasmonic system are of sufficient quality for cryptographic applications --- without requiring any classical post-processing.  This is a significant improvement compared to a previously reported plasmonic quantum random number generator~\cite{plasmonicbs} and a previously reported on-chip time-of-arrival generator~\cite{chip2}, both of which required a randomness extractor to pass the NIST Statistical Test Suite.  A more elaborate list of comparisons with previously reported work in the domain of time-of-arrival based quantum random number generation is presented in \tabref{tab:comparison}.

\begin{table}[]
\begin{tabular}{|l|l|l|l|}
\hline
\textbf{Test}              & \textbf{Req} & \textbf{Prop} & \textbf{p-value} \\ \hline
Frequency                  & 783          & 792           & 0.634516         \\ \hline
Block Frequency            & 783          & 792           & 0.138267         \\ \hline
Cumulative Sums 1          & 783          & 789           & 0.928563         \\ \hline
Cumulative Sums 2          & 783          & 796           & 0.482223         \\ \hline
Runs                       & 783          & 794           & 0.739918         \\ \hline
Longest Run of Ones        & 783          & 795           & 0.894201         \\ \hline
Binary Matrix Rank         & 783          & 793           & 0.757297         \\ \hline
Discrete Fourier Transform & 783          & 793           & 0.021262         \\ \hline
Non-overlapping Template*  & 783          & 792           & 0.573621         \\ \hline
Overlapping Template       & 783          & 788           & 0.805107         \\ \hline
Universal Statistical      & 783          & 795           & 0.487074         \\ \hline
Approximate Entropy        & 783          & 790           & 0.562080         \\ \hline
Random Excursions*         & 472          & 479.5         & 0.472780         \\ \hline
Random Excursions Variant* & 472          & 481           & 0.540922         \\ \hline
Serial 1                   & 783          & 796           & 0.934318         \\ \hline
Serial 2                   & 783          & 793           & 0.219006         \\ \hline
Linear Complexity          & 783          & 794           & 0.444226         \\ \hline
\end{tabular}
\caption{NIST Statistical Test Suite results for the generated sample.  `Req' shows the minimum number of 800 sequences which need to pass a test for the sample to pass the test.  `Prop' shows the number of sequences of the generated sample which passed each test.  For tests which involve more than five subtests (marked with *) the median of the results is presented.}
\label{tab:NISTgen}
\end{table}

\begin{table*}[]
\begin{tabular}{|l|l|l|l|l|}
\hline
\textbf{Work}                              & \textbf{Speed (in Mbits/s)} & \textbf{NIST Tests} & \textbf{Randomness Extractor} & \textbf{On-Chip} \\ \hline
Dynes \textit{et al.}~\cite{toa3}          & 4.01                        & Passed              & None                          & No               \\ \hline
Nie \textit{et al.}~\cite{toa6}            & 96                          & Passed              & Toeplitz Matrix Hashing       & No               \\ \hline
Banerjee \textit{et al.}~\cite{toa8}       & 2.4                         & Passed              & None                          & No               \\ \hline
Khanmohammadi \textit{et al.}~\cite{chip2} & 1                           & Passed              & XOR Hashing                   & Yes              \\ \hline
Current Work                               & 14.4                        & Passed              & None                          & Partially        \\ \hline
\end{tabular}
\caption{Comparison of time-of-arrival based quantum random number generators.}
\label{tab:comparison}
\end{table*}

\begin{figure}
    \centering
    \begin{subfigure}[b]{.48\textwidth}
        \centering
        \includegraphics[scale=0.5]{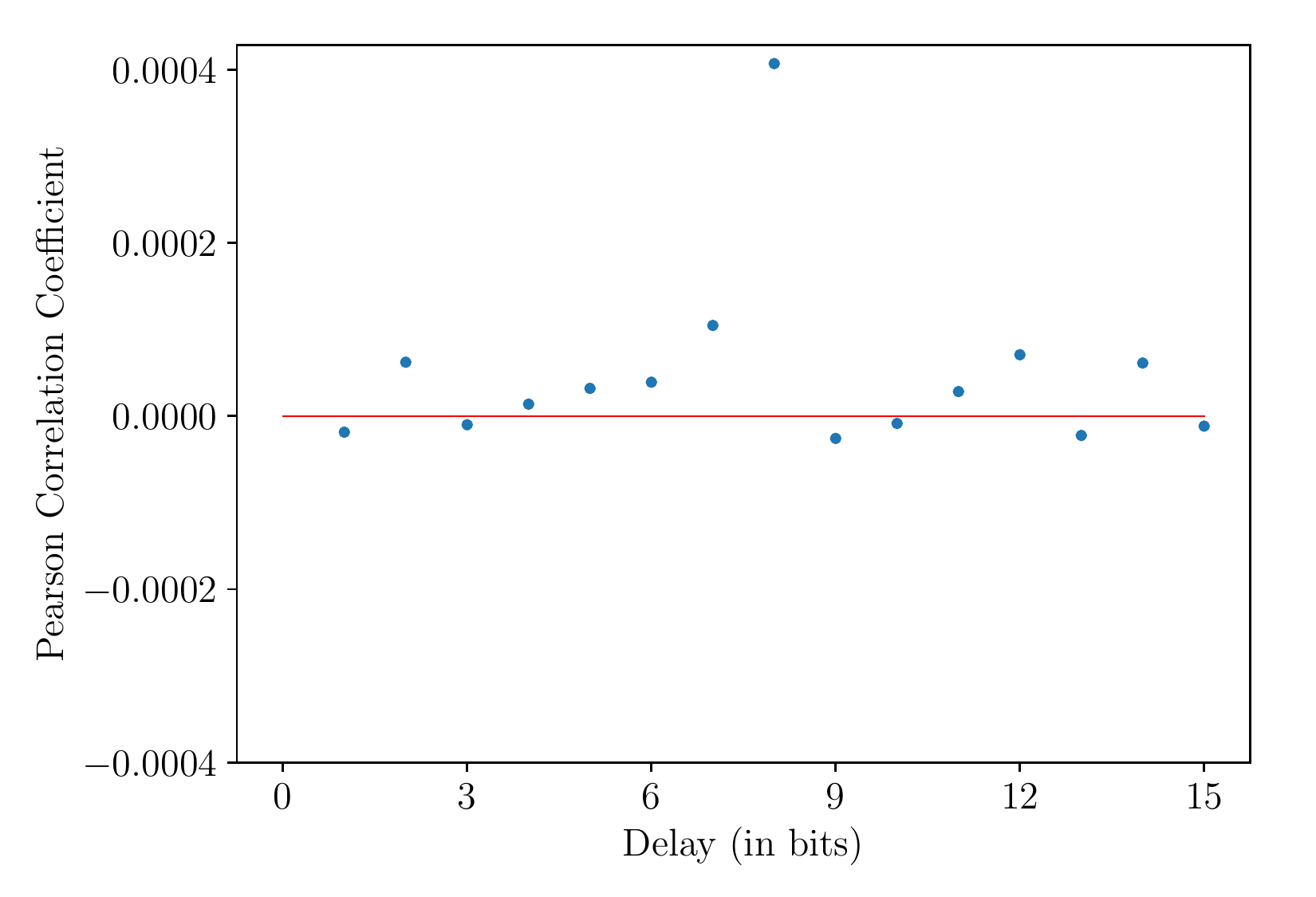}
        \caption{raw sample}
        \label{fig:pearsonraw}
    \end{subfigure}
    
    \vspace{0.25cm}
    
    \begin{subfigure}[b]{.48\textwidth}
        \centering
        \includegraphics[scale=0.5]{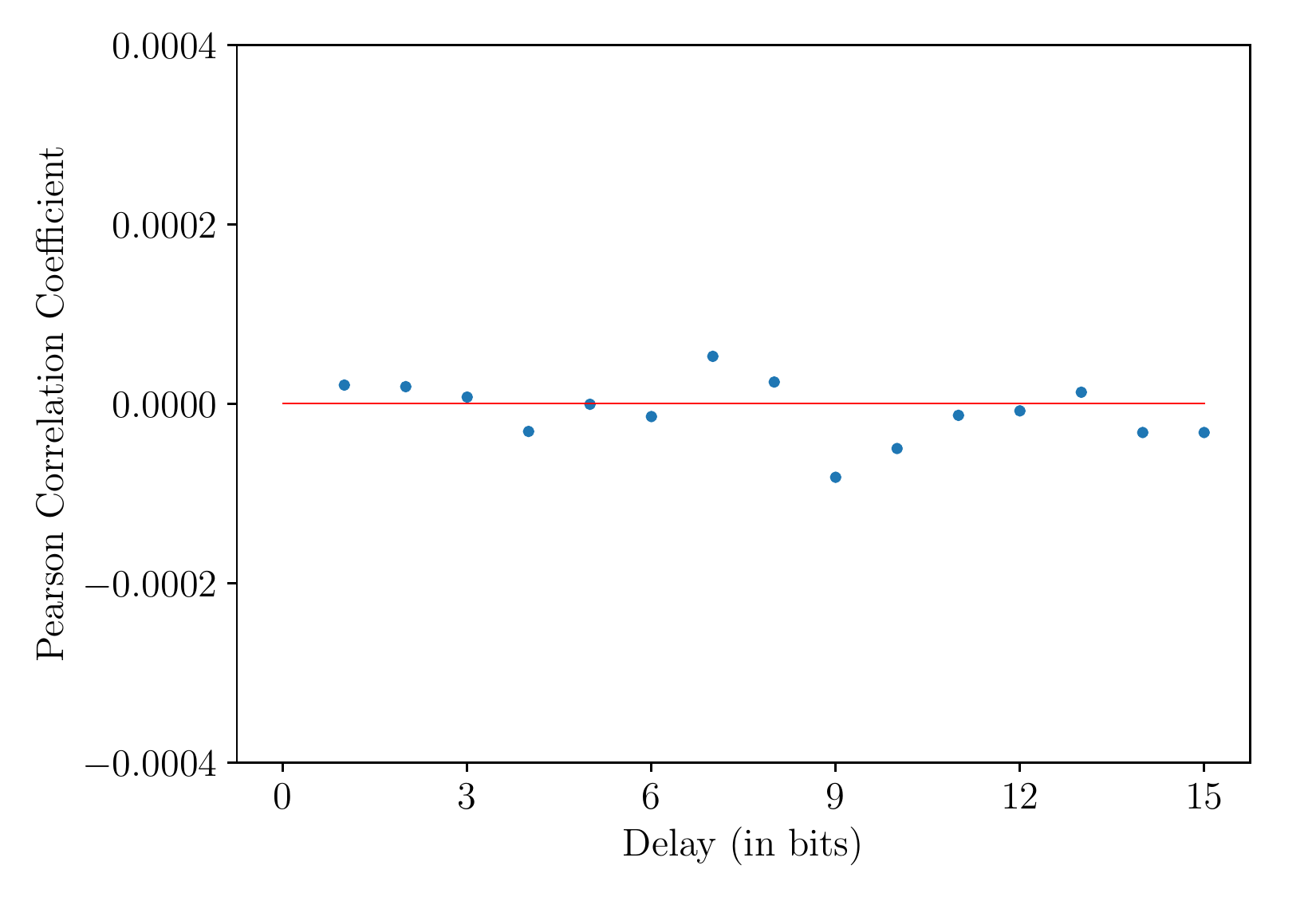}
        \caption{shuffled sample}
        \label{fig:pearsonshuf}
    \end{subfigure}
    \caption{Pearson correlation coefficient of the (a)~raw sample and (b)~shuffled sample with $1$-bit to $15$-bit delays of itself.}
    \label{fig:pearsonrawshuf}
\end{figure}

\begin{table*}[]
\begin{tabular}{|l|l|l|l|}
\hline
\textbf{Test}                  & \textbf{Raw}              & \textbf{Shuffled}         & \textbf{Expected}           \\ \hline
Entropy                        & 7.999997                  & 7.999998                  & 8.000000                    \\ \hline
$\chi^2$ Distribution          & $<$0.01\%                 & 32.17\%                   & 10--90\%                    \\ \hline
Arithmetic Mean                & 127.510                   & 127.524                   & 127.500                     \\ \hline
Monte Carlo value for $\pi$    & 3.14142925                & 3.14063437                & 3.14159265                  \\ \hline
Serial Correlation Coefficient & 0.004634                  & $-$0.000002               & 0.000000                    \\ \hline
\end{tabular}
\caption{ENT Statistical Test Suite results for the raw sample and the shuffled sample.  `Raw' shows the values obtained using the raw sample.  `Shuffled' shows the values obtained using the shuffled sample.  `Expected' shows the expected values for a true random sample.}
\label{tab:ENTrawshuf}
\end{table*}

\begin{table*}[]
\begin{tabular}{|l|l|ll|ll|}
\hline
\textbf{}                  & \textbf{}    & \multicolumn{2}{l|}{\textbf{Raw Sample}}              & \multicolumn{2}{l|}{\textbf{Shuffled Sample}}         \\ \cline{3-6} 
\textbf{Test}              & \textbf{Req} & \multicolumn{1}{l|}{\textbf{Prop}} & \textbf{p-value} & \multicolumn{1}{l|}{\textbf{Prop}} & \textbf{p-value} \\ \hline
Frequency                  & 783          & \multicolumn{1}{l|}{794}           & 0.465415         & \multicolumn{1}{l|}{794}           & 0.465415         \\ \hline
Block Frequency            & 783          & \multicolumn{1}{l|}{789}           & 0.028577         & \multicolumn{1}{l|}{794}           & 0.134365         \\ \hline
Cumulative Sums 1          & 783          & \multicolumn{1}{l|}{793}           & 0.598138         & \multicolumn{1}{l|}{794}           & 0.394631         \\ \hline
Cumulative Sums 2          & 783          & \multicolumn{1}{l|}{795}           & 0.629311         & \multicolumn{1}{l|}{795}           & 0.701879         \\ \hline
Runs                       & 783          & \multicolumn{1}{l|}{791}           & 0.734904         & \multicolumn{1}{l|}{794}           & 0.346453         \\ \hline
Longest Run of Ones        & 783          & \multicolumn{1}{l|}{790}           & 0.798139         & \multicolumn{1}{l|}{786}           & 0.816537         \\ \hline
Binary Matrix Rank         & 783          & \multicolumn{1}{l|}{793}           & 0.964295         & \multicolumn{1}{l|}{788}           & 0.759756         \\ \hline
Discrete Fourier Transform & 783          & \multicolumn{1}{l|}{793}           & 0.052778         & \multicolumn{1}{l|}{791}           & 0.455937         \\ \hline
Non-overlapping Template*  & 783          & \multicolumn{1}{l|}{792}           & 0.525357         & \multicolumn{1}{l|}{792}           & 0.550606         \\ \hline
Overlapping Template       & 783          & \multicolumn{1}{l|}{795}           & 0.373203         & \multicolumn{1}{l|}{784}           & 0.729870         \\ \hline
Universal Statistical      & 783          & \multicolumn{1}{l|}{788}           & 0.130557         & \multicolumn{1}{l|}{795}           & 0.053627         \\ \hline
Approximate Entropy        & 783          & \multicolumn{1}{l|}{792}           & 0.379555         & \multicolumn{1}{l|}{792}           & 0.549331         \\ \hline
Random Excursions*         & 483          & \multicolumn{1}{l|}{490}           & 0.538512         & \multicolumn{1}{l|}{489}           & 0.597986         \\ \hline
Random Excursions Variant* & 483          & \multicolumn{1}{l|}{491}           & 0.209902         & \multicolumn{1}{l|}{491}           & 0.314481         \\ \hline
Serial 1                   & 783          & \multicolumn{1}{l|}{786}           & 0.196920         & \multicolumn{1}{l|}{795}           & 0.587791         \\ \hline
Serial 2                   & 783          & \multicolumn{1}{l|}{792}           & 0.722284         & \multicolumn{1}{l|}{794}           & 0.467799         \\ \hline
Linear Complexity          & 783          & \multicolumn{1}{l|}{788}           & 0.324821         & \multicolumn{1}{l|}{795}           & 0.737414         \\ \hline
\end{tabular}
\caption{NIST Statistical Test Suite results for the raw sample and the shuffled sample.  `Req' shows the minimum number of 800 sequences which need to pass a test for the samples to pass the test.  `Prop' shows the number of sequences of the raw sample or the shuffled sample which passed each test.  For tests which involve more than five subtests (marked with *) the median of the results is presented.}
\label{tab:NISTrawshuf}
\end{table*}

In principle, the random number generation rate can be increased by increasing the light intensity, which increases the photon detection rate.  However, with an increased photon detection rate, the detector dead time becomes non-negligible and the photon counts obtained in an experiment would typically need to be multiplied by a non-unit correction factor to compensate for the resulting underestimation of photon counts.  In our experiment, the result is photon detection times, not photon counts, and so we are unable to correct for undetected photons.  We therefore investigate the effect of detection with a non-unit correction factor on the quality of the random numbers generated in an experiment.  To this end, we adjust the light intensity so as to give a photon detection rate of $5.2\,\text{Mcounts/s}$.  In \appendref{append:correction}, we show that despite the increased photon detection rate, higher order photon number within a time interval of length $T$ is negligible.  However, with an increased detection rate, the average time interval between photon detections decreases to $0.19\,\mu\text{s}$, which is closer to the dead time of the SPAD detector.  The correction factor is about $1.143$ (see \appendref{append:correction}), which means that on average, one in every seven photons arriving at the SPAD detector are not detected.  Furthermore, increasing the photon detection rate to $5.2\,\text{Mcounts/s}$ increases the mean number of photons in a time interval of length $T$ to $0.076$, which slightly decreases the min-entropy to $0.993$ per bit (see \appendref{append:entropy}).  Nevertheless, the min-entropy of the raw bits generated by our setup is still very close to the information-theoretically optimal value of $1$ per bit and so we would require minimal, if any, classical post-processing.  We generated $1,242,469,056$ bits in $30\,\text{s}$, which corresponds to a random number generation rate of about $41.4\,\text{Mbits/s}$.  We then applied the same industry standard tests to the first $800\,\text{Mbits}$ generated, which we will refer to as the raw sample.

We find that short-ranged correlations in the raw sample are mostly negligible, with non-negligible correlations present only between bits at $8$-bit intervals (see \figref{fig:pearsonraw}).  To remove these correlations, we deterministically rearrange or shuffle the bits in the raw sample.  In what follows, we will refer to the resulting sample of bits as the shuffled sample.  As can be seen in \figref{fig:pearsonshuf}, shuffling the bits in the raw sample indeed removed the $8$-bit interval correlations.  The relative frequency of zeros and ones is 0.49993 and 0.50007 respectively, in both the raw sample and the shuffled sample, and so the bias is negligible in both samples.

The ENT Statistical Test Suite results for the raw sample and the shuffled sample are given in \tabref{tab:ENTrawshuf}.  For the raw sample, the $\chi^2$ distribution and the serial correlation coefficient deviate significantly from the expected values for a true random sample.  The large serial correlation coefficient for the raw sample seems to suggest that undetected photons result in correlations between adjacent bytes extracted from consecutive photon arrival times.  In contrast, the $\chi^2$ distribution and the serial correlation coefficient obtained using the shuffled sample show excellent agreement with the expected values for a true random sample.  The negligible serial correlation coefficient for the shuffled sample shows that rearranging the bits in the raw sample, so that the eight consecutive bits which make up a given byte are extracted from eight different non-consecutive photon arrival times, removes the correlations between adjacent bytes.

Finally, we note that both the raw sample and the shuffled sample passed the NIST Statistical Test Suite (see \tabref{tab:NISTrawshuf}).  Hence both these samples are of sufficient quality for cryptographic applications.  This implies that the short-ranged correlations in the raw sample, while non-negligible, are too small to impair the quality to such an extent that the raw sample is unusable for cryptographic applications.  This confirms that a deterministic shuffle, which essentially just transforms short-ranged correlations into long-ranged correlations, is sufficient to improve the quality of the raw sample and more sophisticated randomness extraction schemes are not needed.  In summary, we were able to show that performing detection with a non-unit correction factor introduces small short-ranged correlations into the random bits generated by our plasmonic system, which can be removed with a simple deterministic shuffle provided that the correction factor is not too large.

%%%%%%%%%%%%%%%%%%%%%%%%%%%%
%%%%%%%%%%%%%%%%%%%%%%%%%%%%
%%%%%%%%%%%%%%%%%%%%%%%%%%%%
%%%%%%%%%%%%%%%%%%%%%%%%%%%%
\section{Conclusion}\label{sec:conclusion} 

Our work demonstrates the successful integration of a nanowire plasmonic waveguide into an optical time-of-arrival based quantum random number generation setup.  The specific advantage offered by introducing an on-chip nanowire plasmonic waveguide into the setup is that light in the plasmonic waveguide can be confined to a length scale well below the diffraction limit, which enables the footprint of the on-chip waveguide to be reduced to a size unattainable with dielectric hardware~\cite{plasmonicsdl1, plasmonicsdl2}.  Despite the presence of loss in the plasmonic waveguide and in the optical setup, we first managed to achieve a random number generation rate of $14.4\,\text{Mbits/s}$.  This is an order of magnitude improvement in speed compared to both a previous plasmonic quantum random number generator~\cite{plasmonicbs} and a previous on-chip time-of-arrival generator~\cite{chip2}.  Furthermore, unlike these previous devices, our generator did not require any classical post-processing to pass the NIST Statistical Test Suite.  We were able to increase the generation rate to $41.4\,\text{Mbits/s}$, with the resulting bits only requiring a shuffle to pass all the tests.  Our study makes an important contribution to addressing the on-going challenge of miniaturising on-chip quantum random number generators, as it shows how an on-chip nanoscale plasmonic component, with a footprint well below that of equivalent state-of-the-art dielectric components, can be successfully employed in quantum random number generation.

We note that although our current setup relies on an off-chip source and the detection is also done off-chip, future work on the integration of an on-chip source~\cite{source1, source2, source3, source4} and detector~\cite{detector1, detector2, detector3, detector4} would enable a self-contained plasmonic quantum random number generator chip with a footprint an order of magnitude smaller than its dielectric counterpart.  The primary challenge associated with integrating an on-chip source into our nanowire plasmonic waveguide is that the on-chip near-field single-photon sources typically used in plasmonic systems emit light of a much lower intensity than that of the off-chip source currently used in our setup~\cite{source2, source3, source4}.  A reduction in the light intensity would result in a reduction in the random number generation rate.  Thus, the development of a bright on-chip light source will be key in a future integrated version of our device.  On the other hand, the primary challenge associated with integrating an on-chip detector into our nanowire plasmonic waveguide is that the on-chip near-field superconducting single-photon detectors typically used in plasmonic systems require cryogenic cooling to around $4\,\text{K}$~\cite{detector3, detector4}.  This would greatly increase the footprint and power consumption of the overall system.  Thus, the development of compact cryogenic cooling systems will be important in a future integrated version of our device.  Nevertheless, if these challenges can be overcome, then the successful demonstration of plasmonic quantum random number generation with a fully integrated on-chip near-field source and detector would lead to new opportunities in compact and scalable quantum random number generation.

%%%%%%%%%%%%%%%%%%%%%%%%%%%%
%%%%%%%%%%%%%%%%%%%%%%%%%%%%
%%%%%%%%%%%%%%%%%%%%%%%%%%%%
%%%%%%%%%%%%%%%%%%%%%%%%%%%%
\section*{Data availability statement}

The data that support the findings of this study are available upon reasonable request from the authors.

\begin{acknowledgements}
This research was supported by the South African National Research Foundation, the Harry Crossley Foundation, the University of Stellenbosch, the South African Research Chair Initiative of the Department of Science and Technology (DSI) and National Research Foundation and the DSI South African Quantum Technology Initiative (SAQuTI).  S.K.O.~acknowledges the support from the Air Force Office of Scientific Research (AFOSR) Multidisciplinary University Research Initiative (MURI) Award on Programmable systems with non-Hermitian quantum dynamics (Award No.~FA9550-21-1-0202) and the AFOSR Award FA9550-18-1-0235.
\end{acknowledgements}

%%%%%%%%%%%%%%%%%%%%%%%%%%%%
%%%%%%%%%%%%%%%%%%%%%%%%%%%%
%%%%%%%%%%%%%%%%%%%%%%%%%%%%
%%%%%%%%%%%%%%%%%%%%%%%%%%%%

\appendix

\onecolumngrid

\newpage

%%%%%%%%%%%%%%%%%%%%%%%%%%%%
%%%%%%%%%%%%%%%%%%%%%%%%%%%%
%%%%%%%%%%%%%%%%%%%%%%%%%%%%
%%%%%%%%%%%%%%%%%%%%%%%%%%%%
\section{COMSOL simulation of the nanowire plasmonic waveguide mode}\label{append:simulation} 

The effective mode index of the characteristic mode of the on-chip gold nanowire plasmonic waveguide used in the experiments was determined numerically by 2D finite element method simulation in COMSOL.  We found that $n_{\text{eff}}=1.84+0.0573\text{i}$ for a vacuum wavelength of $785\,\text{nm}$.  For the purpose of the 2D simulation, a cross-section of the nanowire waveguide was modelled as a gold square with a side length of $70\,\text{nm}$, such that one side is adjacent to a $7\,\mu\text{m} \times 3.5\,\mu\text{m}$ silica glass substrate and the other three sides are surrounded by a $7\,\mu\text{m} \times 3.5\,\mu\text{m}$ region of air.  Scattering boundary conditions were enforced at the outer boundaries of the silica glass substrate and the region of air, so that any scattered plane waves were transmitted through these outer boundaries and the materials surrounding the gold nanowire waveguide were modelled as being infinite in size.  The refractive index of gold was estimated using the Lorentz–Drude model proposed by Raki\'{c} \textit{et al.}~\cite{goldLD}.

\begin{figure}[h!]
    \centering
    \begin{subfigure}[b]{.48\textwidth}
        \centering
        \includegraphics[scale=0.5]{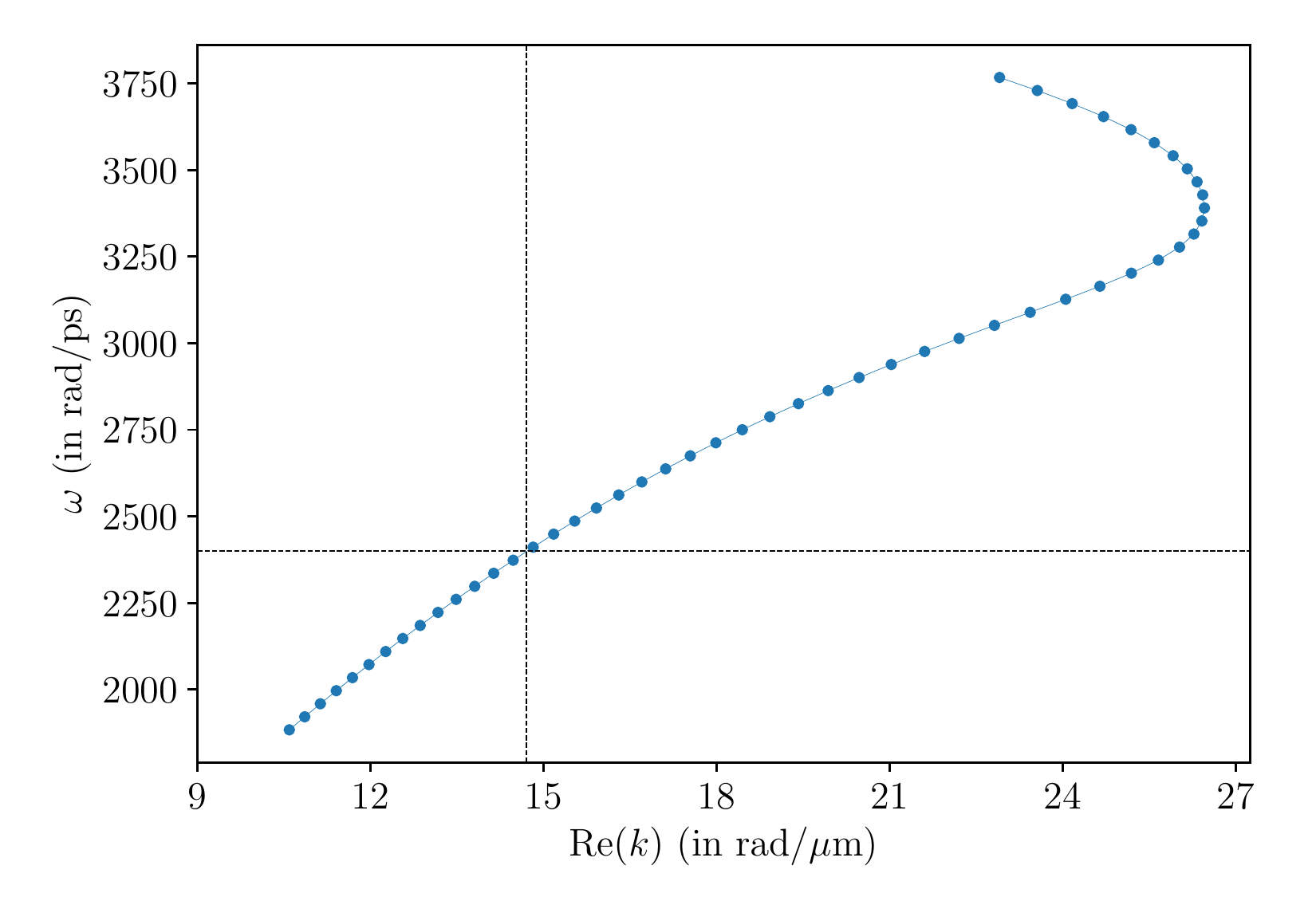}
        \caption{Propagation}
        \label{fig:propagation}
    \end{subfigure}
    \begin{subfigure}[b]{.48\textwidth}
        \centering
        \includegraphics[scale=0.5]{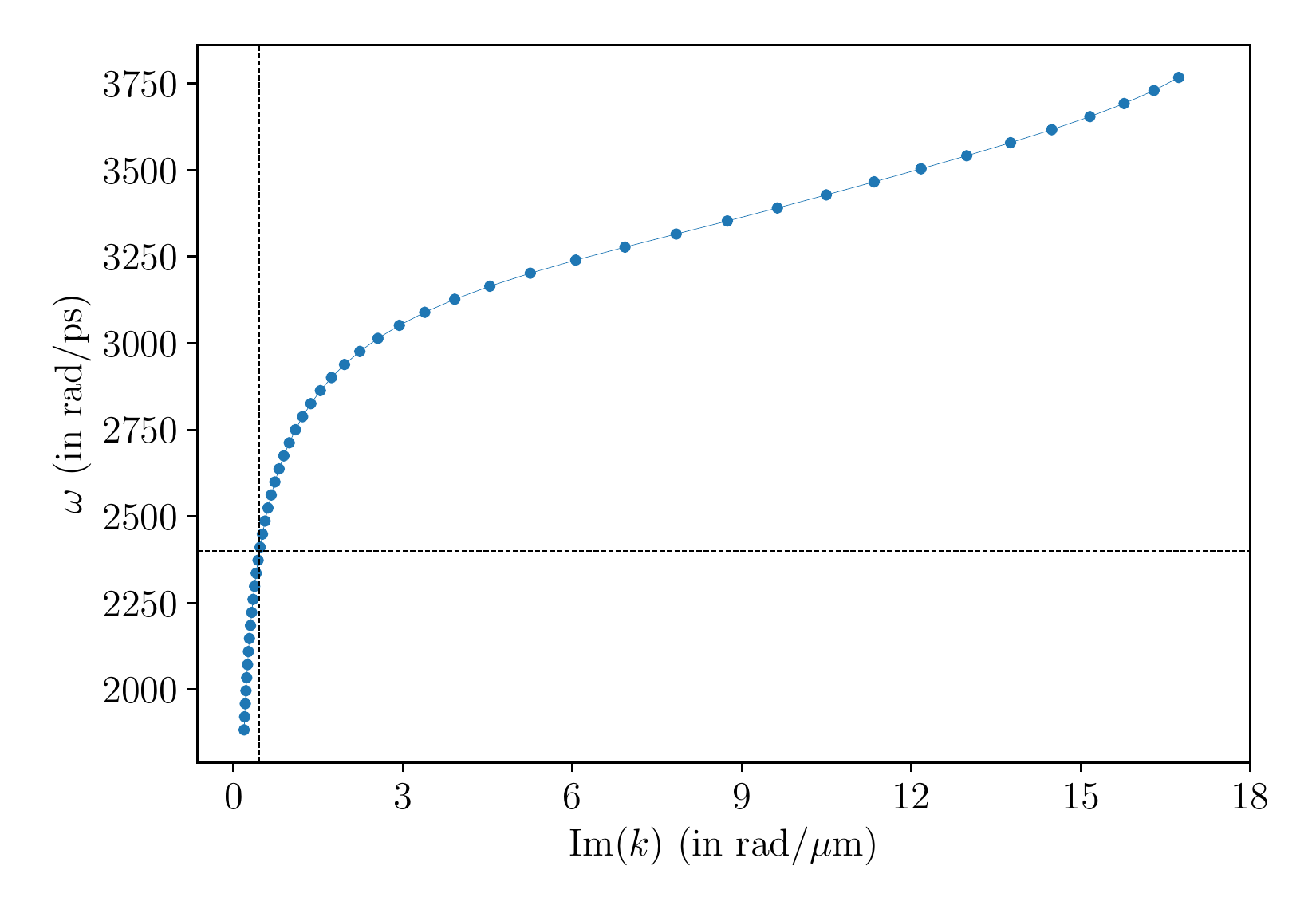}
        \caption{Attenuation}
        \label{fig:attenuation}
    \end{subfigure}
    \caption{Dispersion relation for the on-chip gold nanowire plasmonic waveguide used in the experiments.  (a)~Propagation shows the angular frequency versus propagation constant $\text{Re}(k)$.  (b)~Attenuation shows the angular frequency versus attenuation constant $\text{Im}(k)$.}
    \label{fig:dispersion}
\end{figure}

For further analysis of the waveguide mode, we also determined the propagation and attenuation constants for vacuum wavelengths in the range $500\,\text{nm}$ to $1000\,\text{nm}$.  The resulting dispersion relation is shown in \figref{fig:dispersion}.  The dispersion relation confirms that for a vacuum wavelength of $785\,\text{nm}$ (horizontal line), the nanowire plasmonic waveguide is indeed operating in the surface plasmon polariton regime~\cite{plasmonicsdl1}.

%%%%%%%%%%%%%%%%%%%%%%%%%%%%
%%%%%%%%%%%%%%%%%%%%%%%%%%%%
%%%%%%%%%%%%%%%%%%%%%%%%%%%%
%%%%%%%%%%%%%%%%%%%%%%%%%%%%
\section{Power transmission factor of the nanowire plasmonic waveguide}\label{append:loss} 

We write the net power transmission factor from the rear aperture of the DLM objective (before entry) to the SPAD detector (see \figref{fig:setup}) as $\eta=\eta_{\text{DLM}}\eta_{\text{wgd}}\eta_{\text{col}}$, where $\eta_{\text{DLM}}$ is the input power transmission factor of the DLM objective, $\eta_{\text{wgd}}$ is the power transmission factor of the nanowire plasmonic waveguide and $\eta_{\text{col}}$ is the power transmission factor of the collection optics (which includes the output power transmission factor of the DLM objective, the knife-edge mirror (KM), the fibre coupler (FC) and the multi-mode optical fibre (MM)).  By calculating the ratio of the power measured at the focal point and rear aperture of the DLM objective, we found that $\eta_{\text{DLM}}=0.94$.  To determine $\eta_{\text{col}}$, we used the MM fibre to connect the FC to the continuous-wave laser.  By calculating the ratio of the power measured at the focal point of the DLM objective and the output power of the laser, we found that $\eta_{\text{col}}=0.30$.  This is under the assumption that loss in the collection optics is symmetric.  Finally, we determined $\eta$ by calculating the ratio of the power at the SPAD detector ($P_{\text{out}}$) and the rear aperture of the DLM objective ($P_{\text{in}}$).  We measured $P_{\text{in}}=0.24\,\mu\text{W}$ and calculated $P_{\text{out}}=\frac{Rhc}{\lambda}=4.6\,\text{pW}$ for a photon detection rate of $R=1.8\,\text{Mcounts/s}$.  Hence $\eta=1.9\times 10^{-5}$.  Combining all of the above results, it follows that $\eta_{\text{wgd}}=6.7\times 10^{-5}$.

We can now use the results of the COMSOL simulation to estimate the net power transmission factor of a grating and tapering region (see \figref{fig:waveguide}) under the assumption that the power transmission factors of the input and output gratings and tapering regions are equal.  To this end, we write the power transmission factor of the nanowire plasmonic waveguide as $\eta_{\text{wgd}}=\eta_{\text{grt}}^{2}\eta_{\text{tpr}}^{2}\eta_{\text{nwr}}$, where $\eta_{\text{grt}}$ is the power transmission factor of a grating, $\eta_{\text{tpr}}$ is the power transmission factor of a tapering region and $\eta_{\text{nwr}}$ is the power transmission factor of the nanowire.  To obtain $\eta_{\text{nwr}}$, we note that the power is proportional to the light intensity, which is in turn proportional to $|\mathbf{E}|^2$, and that $|\mathbf{E}|$ is proportional to $e^{-\text{Im}(k)z}$, and so $\eta_{\text{nwr}}=e^{-2\text{Im}(k)z}=0.06$, where $\text{Im}(k)=\text{Im}(n_{\text{eff}})k_{0}=0.459\,\text{rad/}\mu\text{m}$ is the attenuation constant for a vacuum wavelength of $785\,\text{nm}$ and $z\approx 3\,\mu\text{m}$ is the length of the nanowire.  It follows that $\eta_{\text{grt}}\eta_{\text{tpr}}=0.03$.

%%%%%%%%%%%%%%%%%%%%%%%%%%%%
%%%%%%%%%%%%%%%%%%%%%%%%%%%%
%%%%%%%%%%%%%%%%%%%%%%%%%%%%
%%%%%%%%%%%%%%%%%%%%%%%%%%%%
\section{Polarisation dependence of the photon detection rate}\label{append:polarisation} 

\begin{figure}[h!]
    \centering
    \includegraphics[scale=0.5]{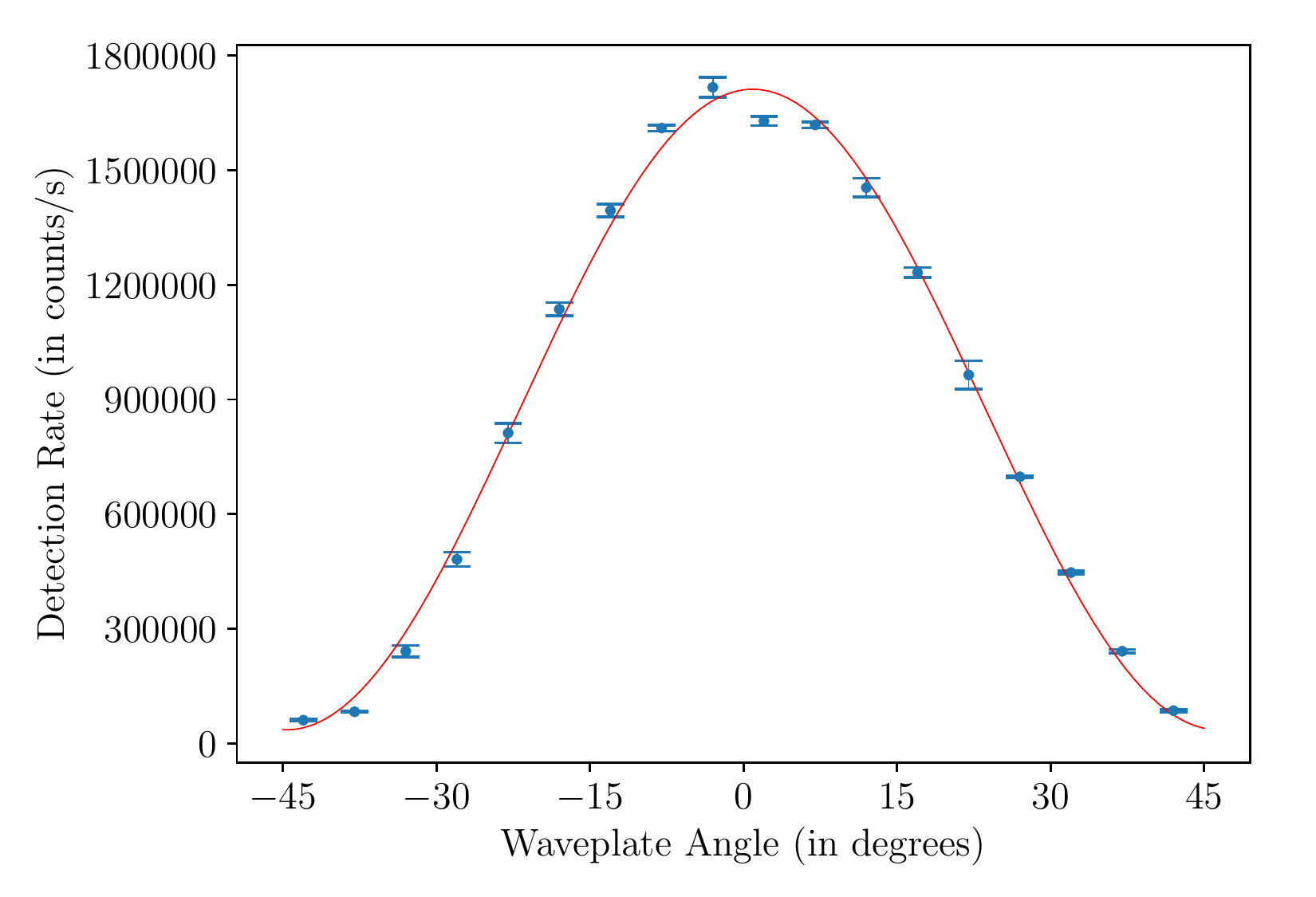}
    \caption{Photon detection rate versus waveplate angle.  Data points are an average of five repetitions and the errors are given by the standard deviation.}
    \label{fig:polarisation}
\end{figure}

We investigate the dependence of the photon detection rate on the polarisation of the input beam.  To this end, we use the second HWP in our experimental setup (see \figref{fig:setup}) to adjust the polarisation of the input beam.  A plot of photon detection rate versus waveplate angle is shown in \figref{fig:polarisation}.  As expected, the photon detection rate has a sinusoidal dependence on the waveplate angle~\cite{polarisation1, polarisation2}.  This confirms that the collection optics is indeed capturing out-coupled photons from the output grating of the nanowire plasmonic waveguide and not scattered photons from the input beam.

%%%%%%%%%%%%%%%%%%%%%%%%%%%%
%%%%%%%%%%%%%%%%%%%%%%%%%%%%
%%%%%%%%%%%%%%%%%%%%%%%%%%%%
%%%%%%%%%%%%%%%%%%%%%%%%%%%%
\section{Proof of uniformity for the time-of-arrival scheme}\label{append:proof} 

We note that for laser light, the probability of $k$ photons arriving in a time interval of length $t$ is given by $P_{t}(k)=e^{-\lambda t}\frac{(\lambda t)^k}{k!}$~\cite{Loudon}, where $\overline{k_{t}}=\lambda t$ is the mean number of photons in a time interval of length $t$ and $\lambda$ is the time-independent mean photon flux.

Consider a time interval of length $T$ divided into two equal sections of duration $\tau_{1}=\tau_{2}=\tau=\frac{T}{2}$.  Let $y$ represent the case that one photon arrives in the time interval of length $T$ (with probability $p(y)$) and let $x$ represent the case that one photon arrives in the first section $[0, \tau_{1}]$ and not in the second section (with probability $p(x)$).  The probability that one photon arrives in the first section given that one photon arrives in the time interval of length $T$ is
\begin{equation}
p_{1}=p(x\,|\,y)=\frac{p(x \land y)}{p(y)}=\frac{p(y\,|\,x)p(x)}{p(y)}.
\end{equation}
We then have that $p(y\,|\,x)=1$, $p(x)=P_{\tau_{1}}(1)P_{\tau_{2}}(0)$ and $p(y)=P_{T}(1)$, which gives
\begin{equation}
p_{1}=\frac{P_{\tau_{1}}(1)P_{\tau_{2}}(0)}{P_{T}(1)}=\frac{e^{-\lambda\tau}\lambda\tau e^{-\lambda\tau}}{e^{-\lambda T}\lambda T}=\frac{\tau}{T}=\frac{1}{2},
\end{equation}
where we have used $T=2\tau$.  Similarly we have the probability that one photon arrives in the second section $[\tau_{1},T]$ and not in the first section as
\begin{equation}
p_{2}=\frac{P_{\tau_{2}}(1)P_{\tau_{1}}(0)}{P_{T}(1)}=\frac{1}{2}.
\end{equation}

These results extend naturally to the more general case where the time interval of length $T$ is divided into $N$ equal sections of duration $\tau_{i}=\tau=\frac{T}{N}$ for $i=1,\ldots,N$.  In particular,
\begin{equation}
p_{i}=\frac{P_{\tau_{i}}(1)P_{(i-1)\tau}(0)P_{(N-i)\tau}(0)}{P_{T}(1)}=\frac{e^{-\lambda\tau}\lambda\tau e^{-\lambda(N-1)\tau}}{e^{-\lambda T}\lambda T}=\frac{\tau}{T}=\frac{1}{N}.
\end{equation}

The above can be generalised to $k$ photons in a time interval of length $T$ with $N$ sections and we have the probability that at least one photon arrives in section $i$ given that no photons occurred before that section as
\begin{align*}
p_{i}&=\frac{\left(\sum\limits_{m=2}^{k}\frac{P_{\tau}(m)}{P_{\tau}(1)}P_{(N-i)\tau}(k-m)+P_{(N-i)\tau}(k-1)\right)P_{\tau}(1)}{P_{T}(k)}P_{(i-1)\tau}(0) \\
&=\frac{1}{P_{T}(k)}\sum_{m=1}^{k}\frac{P_{\tau}(m)}{P_{\tau}(1)}P_{(N-i)\tau}(k-m)P_{\tau}(1)P_{(i-1)\tau}(0).
\end{align*}
Substituting in the $P_{k}(j)$ and using $0^0=1$ as convention, one finds
\begin{align*}
p_{i}&=\sum_{m=1}^{k}\frac{c(k,m)(N-i)^{k-m}}{N^{k}} \\
&=\left(1-\frac{i-1}{N}\right)^{k} - \left(1-\frac{i}{N}\right)^{k},
\end{align*}
which is the form given in Ref.~\cite{toa6}.

%%%%%%%%%%%%%%%%%%%%%%%%%%%%
%%%%%%%%%%%%%%%%%%%%%%%%%%%%
%%%%%%%%%%%%%%%%%%%%%%%%%%%%
%%%%%%%%%%%%%%%%%%%%%%%%%%%%
\section{Mean photon number and min-entropy estimation}\label{append:entropy} 

For laser light, the probability of $k$ photons arriving in a time interval of length $t$ is given by $P_{t}(k)=e^{-\lambda t}\frac{(\lambda t)^k}{k!}$~\cite{Loudon}, where $\overline{k_{t}}=\lambda t$ is the mean number of photons in a time interval of length $t$ and $\lambda$ is the time-independent mean photon flux.

For a photon detection rate $R$, there are $R$ dead times within one second, each of which is $\tau_{d}$ in duration, where $\tau_{d}$ is the detector dead time.  The total number of undetected photons in one second is therefore $R\lambda\tau_{d}$, and the total number of detected and undetected photons is $R(1+\lambda\tau_{d})$.  Hence we have that the mean number of photons for one second is
\begin{equation}
\overline{k_{1}}=\frac{\overline{k_{T}}}{T}=R(1+\lambda\tau_{d}),
\end{equation}
which gives $\overline{k_{T}}=RT+R\lambda T\tau_{d}$.  Substituting in $\lambda=\frac{\overline{k_{T}}}{T}$ on the right hand side of the previous equation and rearranging gives
\begin{equation}
\overline{k_{T}}=R\frac{T}{(1-R\tau_{d})}.
\end{equation}
For $R=1.8\,\text{Mcounts/s}$, $T=12.8\,\text{ns}$ and $\tau_{d}=24\,\text{ns}$, we find that $\overline{k_{T}}=0.024$.

The mean photon number $\overline{k_{T}}$ can be used to obtain a conservative estimate of the min-entropy of the bits generated by our experimental setup.  In particular,
\begin{equation}
H_{\text{min}}=\log_2(N)+\log_2(1-e^{-\overline{k_{T}}})-\log_2(\overline{k_{T}}),
\end{equation}
is a lower bound estimate of the min-entropy $H_{\text{min}}$~\cite{toa6}, where $N$ is the total number of bins in a time interval of length $T$.  Substituting in $N=256$ and $\overline{k_{T}}=0.024$ we obtain $H_{\text{min}}=7.98$ per byte or equivalently $H_{\text{min}}=0.998$ per bit.

Furthermore, for an increased photon detection rate of $R=5.2\,\text{Mcounts/s}$, the mean photon number increases to $\overline{k_{T}}=0.076$ and the min-entropy decreases to $H_{\text{min}}=7.95$ per byte or equivalently $H_{\text{min}}=0.993$ per bit.

%%%%%%%%%%%%%%%%%%%%%%%%%%%%
%%%%%%%%%%%%%%%%%%%%%%%%%%%%
%%%%%%%%%%%%%%%%%%%%%%%%%%%%
%%%%%%%%%%%%%%%%%%%%%%%%%%%%
\section{Higher order photon events and correction factor}\label{append:correction} 

For laser light, the probability of $k$ photons arriving in a time interval of length $t$ is given by $P_{t}(k)=e^{-\lambda t}\frac{(\lambda t)^k}{k!}$~\cite{Loudon}, where $\overline{k_{t}}=\lambda t$ is the mean number of photons in a time interval of length $t$ and $\lambda$ is the time-independent mean photon flux.

In \appendref{append:entropy}, we showed that the mean number of photons in a time interval of length $T$ can be written as
\begin{equation}
\overline{k_{T}}=R\frac{T}{(1-R\tau_{d})},
\end{equation}
where $R$ is the photon detection rate and $\tau_{d}$ is the detector dead time.  Using $R=5.2\,\text{Mcounts/s}$, $T=12.8\,\text{ns}$ and $\tau_{d}=24\,\text{ns}$ we obtain $\overline{k_{T}}=0.076$.  We then have that $\lambda=\frac{\overline{k_{T}}}{T}$, and using this we find that the relative probability of a single photon in a time interval of length $T$ compared to the case of higher order photon number is $p_{1}=\frac{P_{T}(1)}{1-P_{T}(0)}=0.96$ and for two photons $p_{2}=\frac{P_{T}(2)}{1-P_{T}(0)}=0.037$.  Thus higher order photon number within a time interval of length $T$ is negligible.

However, undetected photons may introduce some correlations in the random numbers generated.  A correction factor takes into account the non-negligible detector dead time and that any counts a detector measures is an underestimate of the true counts.  The correction factor is the ratio of total photons, which we explain in \appendref{append:entropy} is $R(1+\lambda\tau_{d})$, to photons detected $R$ and is given by
\begin{equation}
c_{F}=\frac{R(1+\lambda\tau_{d})}{R}=1+\lambda\tau_{d}.
\end{equation}
Substituting in
\begin{equation}
\lambda=\frac{\overline{k_{T}}}{T}=\frac{RT}{(1-R\tau_{d})}\frac{1}{T}=\frac{R}{(1-R\tau_{d})}
\end{equation}
we get
\begin{equation}
c_{F}=1+\frac{R\tau_{d}}{(1-R\tau_{d})}=(1-R\tau_{d})^{-1}.
\end{equation}
For $R=5.2\,\text{Mcounts/s}$ and $\tau_{d}=24\,\text{ns}$, we have $c_{F}=1.143$.  This means that one in every $(1.143-1)^{-1}=7$ photons arriving at the detector are not detected.

\end{document}